\documentclass[reprint,aps]{revtex4-2}
\usepackage{graphicx,dcolumn,bm,lipsum,color,ulem}
\usepackage{mathtools,amsmath,amssymb}
\usepackage{dashrule}
\long\def\comment#1{}
\begin{document}

\title{Wait-time Distributions for Photoelectric Detection of Light} 
\author{Luis Felipe Morales Bultron, Reeta Vyas, and Surendra Singh}
\affiliation{Department of Physics, University of Arkansas Fayetteville, Arkansas 72701} 
\date{\today} 

\begin{abstract}
Wait-time distributions for the $n$th photo-detection at a detector illuminated by a stationary light beam are studied. Both unconditional measurements, initiated at an arbitrary instant, and conditional measurements, initiated upon a photo-detection, are considered.  Simple analytic expressions are presented for several classical and quantum sources of light and are used to quantify and compare photon sequences generated by them. These distributions can be measured in photon counting experiments and are useful in characterizing and generating photon sequences with prescribed statistics.  Effects of non-unit detection efficiency are also discussed, and curves are presented to illustrate the behavior.
\end{abstract}

\pacs{ 42.50.-p, 42.50.Ar, 42.65.Lm, 42.50.Lc}
\maketitle

\section{\label{sec:level1}Introduction}
Statistical analysis of photon sequences generated by light sources is a fundamental tool for understanding the dynamics of light sources. By comparing the measurements of photoelectric pulse sequence generated by the light incident on a detector with theoretical predictions, we can correlate the photoelectric pulse dynamics to the source dynamics. Thus, a meaningful characterization of statistical properties of light is limited to quantities that can measured experimentally and calculated theoretically\cite{mandelwolf65,bendjaballah,saleh,carmichael89}.  

Most commonly measured and calculated quantities to characterize the statistical properties of light are the unconditional and conditional photo-count distributions and a few low order time-dependent correlations. An example of the latter is the conditional measurements of light intensity, which is related to the two-time intensity correlation function $g^{(2)}(t; t +T)$ of light \cite{carmichael89,glauber63}.  Another quantity  closely related to the two-time intensity correlation function is the conditional wait-time probability density $w_{1}(T|t)$, such that $w_{1}(T|t)dT$ is the probability that an interval $T$ elapses before the first photo-detection is recorded at time $t+T$, given that a photo-detection was recorded at time $t$.  The function $w_{1}(T|t$) is extremely useful in characterizing bunching and anti-bunching properties of photons in a light beam \cite{vyassingh88,vyassingh00}.

A generalization of the wait-time distribution is $w_{n}(T|t)$, the probability density that an interval $T$ elapses before the $n$-th photo-detection is recorded given that the measurement commenced at a photo-detection at time $t$ \cite{bendjaballah,saleh}. Another related distribution is the unconditional waiting time distribution $P_{n}(t,T)$ such that $P_{n}(t,T)dT$ is the probability that a time $T$ elapses before the $n$th detection occurs at time $t+T$ given that the counting commenced at an arbitrary time $t$.  Wait-time distributions provide a more detailed temporal picture of  photoemissions that goes beyond that afforded by $n$th order intensity correlations. Experimental measurement of these distributions requires detectors of high efficiencies; with low efficiency detectors, their measurement reduces essentially to a measurement of intensity correlations.  Additionally, the calculations of wait-time distributions are usually more difficult than those for intensity correlations. For these reasons, relatively few calculations of these distributions have been carried out. However, the availability of near unit detection efficiency detectors and interest in single-photon sources \cite{sipahigil,leifgen,couteau,cheng} have made these distributions relevant again for characterizing light sources. These are the quantities of interest here. 

The paper is organized as follows. Section I introduces the distributions $P_{n}(t,T)$,  $w_{n}(T|t)$ and other quantities of interest that are used to compare photon sequences emitted by different sources in this paper.  Section {II} presents expressions for these quantities for narrow-band thermal or Gaussian light from a laser below threshold \cite{haken70,loudon00,microcavity}. Section {III} presents wait-time distributions for the light from a degenerate parametric oscillator (DPO) operating below threshold. This is well known to be a source of squeezed light that requires a quantum mechanical treatment to account for its statistical properties \cite{drummond,wolinsky,kimble87,vyas95,vyas03,vines06,vyas09,arnab10}.  This is also of interest as a source of conditional single-photon sequence \cite{christ13}.  Section {IV} considers photon sequence generated in resonance fluorescence from a single two-level atom driven by a coherent field. This is another source of quantum mechanical light that requires quantum mechanics for its description \cite{carmichael76,kimble76,arnoldus}. We also discuss the effects of non-unit detection efficiency on these distributions and find that the effect of non-ideal detection efficiency is significantly different for classical and quantum light.  In all cases, exact or approximate but simple analytic expressions for these distributions are presented.   The paper ends with a summary of principal results and conclusions of the paper in Sec. {V}.

\subsection{Wait-time  Distributions}
Consider a stationary beam of quasi-monochromatic light incident on a photodetector of quantum efficiency $\eta\,(0< \eta \leq  1)$. Throughout the paper, we will refer to photon flux (\# photons/sec) associated with the beam as intensity.  For stationary light, the average of the photon-flux operator $\langle \hat I (t)\rangle (=\langle \hat I \rangle)$ and the photoelectric-detection probabilities are independent of the initial time $t$.  The explicit form for the  photon flux operator depends on the source of light. For light generated by optical cavities, it has the form $2\gamma \hat{n}$ where $\hat{n}$ is the occupation number for the source cavity and $2\gamma$ is the energy decay rate for the source cavity \cite{vyassingh89-1}. For a two-level atom it has the form $2\beta\hat{\sigma}$, where $\hat{\sigma}$ is the inversion operator for the atom and $2\beta$ is Einstein-A coefficient for the atomic transition that results in the emission of photons \cite{carmichael89}. Consequently, the unconditional and conditional waiting time distributions  $P_{n}(t,T)\equiv P_{n}(T)$ and  $w_{n}(T|t) \equiv w_{n}(T)$ are given by \cite{bendjaballah,saleh,carmichael89}: 
\begin{align}
P_{n}(T)&= \bigg\langle \mathcal{T}\textrm{:}\,\eta\hat{I}(T)\frac{(\eta \hat{U}(T))^{n-1}}{(n-1!)}e^{-\eta \hat{U}(T)}\textrm{:}\bigg\rangle\,,\label{Pnt1}\\
w_{n}(T)&= \frac{\eta}{\langle \hat{I} \rangle}\bigg\langle \mathcal{T}\textrm{:}\hat{I}(T)\frac{(\eta \hat{U}(T))^{n-1}}{(n-1!)}e^{-\eta \hat{U}(T)}\hat{I}(0)\textrm{:}\bigg\rangle.
\label{wnt1}\end{align}
Here 
 $\mathcal{T}$ and $:(\;):$ denote, respectively, the time and normal ordering of operators, and $\hat{U}(T) = \int_{0}^{T}\hat{I}(t')dt'$ is the integrated intensity.  Wait-time distributions  $P_{n}(T)$ and $w_n(T)$, as well as the photo-count distribution $p(n,T)$ (the probability of recording $n$ photo-counts in interval $T$), can be obtained from the generating function \cite{bendjaballah,saleh,carmichael89}:
\begin{align}
G(s,T)&= \bigg\langle \mathcal{T}\textrm{:} e^{-s \eta \hat U(T) }  \textrm{:}\bigg\rangle \label{Gst}
\end{align}
where $s$ is a dimensionless parameter. Note that $G(1,T)$ is the probability of no photo-count $p(0,T)$ in the interval $T$. The photo-count and the wait-time distributions can be expressed in terms of $G(s,T)$ as 
\begin{align}
p(n,T) &= \frac{(-1)^{n}}{n!}\bigg[\frac{\partial ^{n}}{\partial s^n} G(s,T)\bigg]_{s =1}\,, \label{pnt}\\
P_{n}(T)&= -\frac{(-1)^{n-1}}{(n-1)!}\bigg[\frac{\partial ^{n-1}}{\partial s^{n-1}} \frac{1}{s} \frac{\partial}{\partial T}G(s,T)\bigg]_{s =1}\,\nonumber\\
&=\,-\,\frac{\partial}{\partial T} \bigg(  \sum_{k=0}^{n-1}p(k,T) \bigg), \label{Pnt2}\\
w_{n}(T) &=\frac{1}{\eta \langle \hat{I}\rangle} \frac{(-1)^{n-1}}{(n-1)!}\bigg[\frac{\partial ^{n-1}}{\partial s^{n-1}} \frac{1}{s^{2}}  \frac{\partial^2}{\partial T^2}G(s,T)\bigg]_{s =1},\nonumber\\ 
& = \frac{1}{\eta \langle \hat{I} \rangle}\frac{\partial^{2}}{\partial T^{2}}\bigg(  \sum_{k=0}^{n-1} (n-k)p(k,T) \bigg)\,.
\label{wnt2}
\end{align}
It is important to note that $n$ and $T$ play dual roles in these equations;  $w_n(T)$ and $P_n(T)$ are distributions of $T$ for fixed $n$, whereas $p(n,T)$ is a distribution of $n$ for fixed counting interval $T$. An inspection of Eqs.~\eqref{pnt} - \eqref{wnt2} shows that $P_{n}(T)$ and $w_{n}(T)$ are related by \cite{arnoldus,arnolduswn,arnolduswn2}
\begin{align}
w_{n}(T) = - \frac{1}{\eta \langle \hat{I}\rangle} \frac{\partial }{\partial T}\bigg( \sum_{k=1}^{n}P_{k}(T) \bigg)\,.
\label{wnt3}\end{align}  
This relation can be used to determine $w_{n}(T)$ once the set $P_{n}(T)$ is known. 

A comparison of these distributions for different sources can reveal similarities or differences in photo-emission sequences such as the dominance of small or large wait-times or bunching or anti-bunching of photons. For long wait times they all decay exponentially, whereas for short time (or low detection efficiency $\eta$), the leading terms of these distributions are given by  
\begin{align}
w_n(T)/(\eta\langle I\rangle)& \approx (\eta\langle I\rangle T)^{(n-1)}g^{(n+1)}(0),\label{wnT=0}\\
P_n(T)/(\eta\langle I\rangle )&\approx (\eta\langle I\rangle T)^{(n-1)} g^{(n)}(0), \label{PnT=0}\end{align}
where $g^{(n)}$ is the normalized $n$th  order ($n\ge1$) intensity correlation function \cite{glauber63}. Equations \eqref{wnT=0}-\eqref{PnT=0} imply that all $n\ge2$ distributions  vanish at $T=0$ and that the most probable wait-time for the second photo-detection is nonzero. The  short time dependence implied by Eqs. \eqref{wnT=0} and \eqref{PnT=0} assumes non-vanishing zero-time intensity correlations.  If these correlations vanish, the leading short-time dependence may be different from that given here. We will see an example of this in the discussion of photon sequences generated in single atom resonance fluorescence. 
 
From the experimental perspective, $n=1$ to 3 distributions are the important ones as this paper shows. Distributions beyond these carry little qualitatively new information.  A comparison of unconditional and conditional wait-time distributions for the same $n$ shows that $P_1$ and $w_1$ can differ the most from each other for a given source, the difference being especially significant at short times [Eqs. \eqref{wnT=0}-\eqref{PnT=0}].  For $n\ge2$, the two types of distributions are qualitatively similar. From the short time behavior [\eqref{wnT=0}-\eqref{PnT=0}], we also see that while $P_2$ and $w_1$  are both proportional to $g^{(2)}(0)$, the former vanishes at zero delay, while the latter can be very large as will be seen for the parametric oscillator. Of course, $w_1(0)$  can also vanish as will be seen for single-atom resonance fluorescence.  Finally, for $T\to\infty$, both $P_{n}(T)$ and $w_{n}(T)$ must decay sufficiently fast for them to be normalizable.  With these preliminaries, we are ready to discuss the waiting time distribution for various light sources.
\subsection{Coherent Light}
We begin by summarizing the properties of photon sequence for coherent light, which may be considered to be the output of a well-stabilized single-mode laser operating far above threshold \cite{scullyzubairy}. Coherent light corresponds to a constant flux [$I(t)=\langle I\rangle$] photon sequence so that  $\int_0^T I(t)dt=\langle  {I} \rangle T$. Using this in Eqs.~\eqref{Gst} and \eqref{pnt}, the generating function and the photo-count distribution for coherent light are found to be  \cite{bendjaballah,saleh}
\begin{align}
G(s,T)&=e^{-s\eta \langle\hat{I}\rangle T}\,,\label{gstcoh}\\
p(n,T) &= \frac{(\eta \langle \hat{I} \rangle T)^{n}e^{-\eta \langle  \hat{I} \rangle T}}{n!}\,.\label{pncoh}
\end{align}
The generating function is thus a simple exponential, and the photo-count distribution is a Poisson distribution as expected of a constant rate sequence of independent (uncorrelated) photons. Since successive photoemissions are uncorrelated, the photoelectric measurements of coherent light beginning at a photo-detection (conditioned on a photo-detection)  or beginning at an arbitrary time (unconditional) coincide. Both the conditional and unconditional wait-time distributions are then given by
\begin{align}
w_{n}(T) = \frac{ \eta \langle \hat{I} \rangle  (\eta \langle \hat{I} \rangle T)^{(n-1)} }{(n-1)!}\;e^{-\eta \langle \hat{I} \rangle T}= P_{n}(T).\label{Pncoh}
\end{align}
In particular, $w_{1}(T)= \eta \langle \hat{I} \rangle e^{-\eta \langle \hat{I} \rangle T} = P_{1}(T)$ are simple exponentials with $T=0$ as the most probable wait-time. In contrast, the wait  time distributions for $n \geq 2$ vanish at $T=0$, implying a finite wait-time for the second (or higher order) photo-detection to occur. 

The wait-time distributions \eqref{Pncoh} for coherent light or a Poisson photon sequence have the form of a gamma distribution with shape parameter $n$  and rate parameter $\eta \langle \hat{I} \rangle$ \cite{arfken}.  It follows from the properties of gamma distribution that the mean and variance of the wait-time for the $n$th photo-detection are given by 
\begin{align}
\langle T\rangle_n&=\frac{n}{\eta \langle \hat{I} \rangle}\,,\label{tavcoh}\\
\langle (\varDelta T)^2\rangle_n&=\frac{n}{(\eta \langle \hat{I} \rangle)^2}\,.\label{tvarcoh}
\end{align}
In writing these equations,  we have not distinguished between conditional and unconditional averages because for a Poisson sequence of uncorrelated photons underlying coherent light, the conditional and unconditional averages coincide.  These quantities for a Poisson photon sequence set the references relative to which long or short wait-times for photo-detection or   bunching or antibunching of photons for other photon sequences can be defined.     
\section{Thermal Light}
Thermal light (narrow-band Gaussian light) is an excellent model for filtered light from electrical discharge lamps or the light from a single-mode laser operating below threshold. In this paper, we will consider thermal light as coming from a single-mode laser operating below threshold with mean cavity photon number $\overline{n}$, cavity decay rate $2\gamma $, and average photon-flux (referred to as intensity throughout the paper)  $\langle \hat{I} \rangle =2\gamma \overline{n}$. The electric field amplitude for such a light beam can be modeled as a complex Gaussian random process with zero mean and variance $\bar{n}$ (average modulus squared of the complex field amplitude, also the mean photon number in laser cavity) \cite{haken70}. The generating function of light emitted by the laser below threshold  is then given by  \cite{bedard,jakemanandpike} 
\begin{align}
&G(s,T) = \frac{e^{2\gamma T}}{ \bigg[ \cosh(zT)+\frac{1}{2}\bigg[  \frac{z}{2\gamma} + \frac{2\gamma}{z} \bigg]\sinh(zT) \bigg]}\,, \label{Gth}
\end{align}
where $z^{2} = (2\gamma )^{2}+2s \eta (2\gamma) \langle \hat{I} \rangle =(2\gamma)^2(1+2s\eta\bar{n})$. Using this generating function in Eq.~\eqref{pnt}, $p(n,T)$ can be expressed in terms of modified Bessel functions ~\cite{bedard}, which can be used in Eqs.~\eqref{Pnt2}--\eqref{wnt2} to obtain $P_{n}(T)$ and $w_{n}(T)$ or Eq. \eqref{Gth} can be used directly in Eqs. \eqref{Pnt2} and \eqref{wnt2} to compute these distributions. The resulting expressions, in general,  have complicated algebraic forms. They simplify considerably, however, for small and large mean photon number $\bar{n}$. We consider these separately.
\subsection{Small mean cavity photon number $\bar{n}$} 
For small mean cavity photon number, $\bar{n}\ll1$,  we expand $G(s,T)$ as a power series in $\bar{n}$, using the constraints imposed by Eqs. \eqref{wnT=0}-\eqref{PnT=0} as guide. Then the number of terms that need to be retained depends on $n$ (subscript of wait-time distribution). This procedure leads to the following expressions for unconditional wait-time distributions 
    \begin{align}
P_1(T)&\approx 2\eta\gamma\bar{n}e^{-2\eta\gamma\bar{n} T}\,,\label{P1th0}\\
P_2(T)&\approx 2\eta\gamma\bar{n}e^{-2\eta\gamma\bar{n} T}\frac{\eta\bar{n}}2\left[1+4\gamma  T-e^{-4\gamma T}\right] \,, \label{P2th0}\\
P_3(T)&\approx 2\gamma\bar{n}e^{-2\eta\gamma\bar{n} T}\frac{(\eta\bar{n})^2}4\left[1+8\gamma  T+8(\gamma T)^2\right.\notag\\
&\quad\left.-e^{-4\gamma T}(12\gamma T+1)\right]\,.\label{P3th0}
\end{align}
A similar procedure for the conditional wait-time distribution leads to
\begin{align}
w_1(T)&\approx2\eta\gamma \bar{n}e^{-2\eta\gamma\bar{n} T}\left(1+e^{-4\gamma T}\right)\,,\label{w1th0}\\
w_2(T)&\approx 2\eta\gamma\bar{n}e^{-2\eta\gamma\bar{n} T} \eta\bar{n}\left[1+2\gamma T\right.\notag\\
&\quad\left.+e^{-4\gamma  T}(-1 + 6\gamma T)\right]\,,\label{w2th0}\\
w_3(T)&\approx 2\eta\gamma\bar{n}e^{-2\eta\gamma\bar{n} T} (\eta \bar{n})^2\left[1+3\gamma T+2(\gamma T)^2 \right.\notag\\
&\quad\left.-e^{-4\gamma T}\left(2+3\gamma T-18(\gamma T)^2\right)+e^{-8\gamma  T} \right].\label{w3th0}
\end{align}
These expressions are compared with the exact waiting time distribution in Fig. 1 for small $\bar{n}=0.01$. The solid curves are computed by using  Eq. \eqref{Gth} in Eqs. \eqref{Pnt2} and \eqref{wnt2} and the dashed curves represent small $\bar{n}$ approximations of Eqs. \eqref{P1th0}--\eqref{w3th0}.   It can be seen that the approximate expressions capture the essential features of unconditional and conditional wait-time distributions both for small times and large times.  

$P_1(T)/2\eta\gamma\bar{n}$ always starts out at the value 1 at $T=0$ and monotonically decreases for large times.   $w_1(T)/2\eta\gamma\bar{n}$, on the other hand, starts out at twice this value as $w_1(0)/2\eta\gamma\bar{n}=g^{(2)}(0)$, which for thermal light is known to have twice the value for coherent photons \cite{loudon00}. $w_1(T)$ also displays a prominent narrow peak at short wait-times riding on top of a long exponential, which can be seen more clearly in the expanded view of $w_1(T)$ in the inset in Fig. 1(b). This prominent short wait-time peak in $w_1(T)$ for small $\bar{n}$ results from bunching of photons in thermal light. Recall, that for small $\bar{n}$, the average photo-detection rate $2\gamma\bar{n}$ is small, but immediately following a photo-detection, the detection rate surges to twice the average rate due to photon bunching. The surge in photo-emission rate lasts only about one cavity lifetime and its importance relative to the average photo-emission rate decreases as the mean cavity photon number grows as will be seen in the next subsection. 

Wait-time distributions for $n\ge2$ start at zero, reach a maximum for some nonzero wait-time, which depends on $n$ and decay to zero for large wait-times. We also note that the two types of distributions  $n\ge3$ are similar. It is also worth pointing out that to the best of our knowledge, these expressions for $P_n$ and $w_n$ have not appeared before in the literature although the statistical properties of thermal light have been studied extensively for a long time \cite{saleh,bedard,jakemanandpike}. 

\begin{figure*}[t]
  \centering
  \vskip1in 
\includegraphics[width=0.45\linewidth]{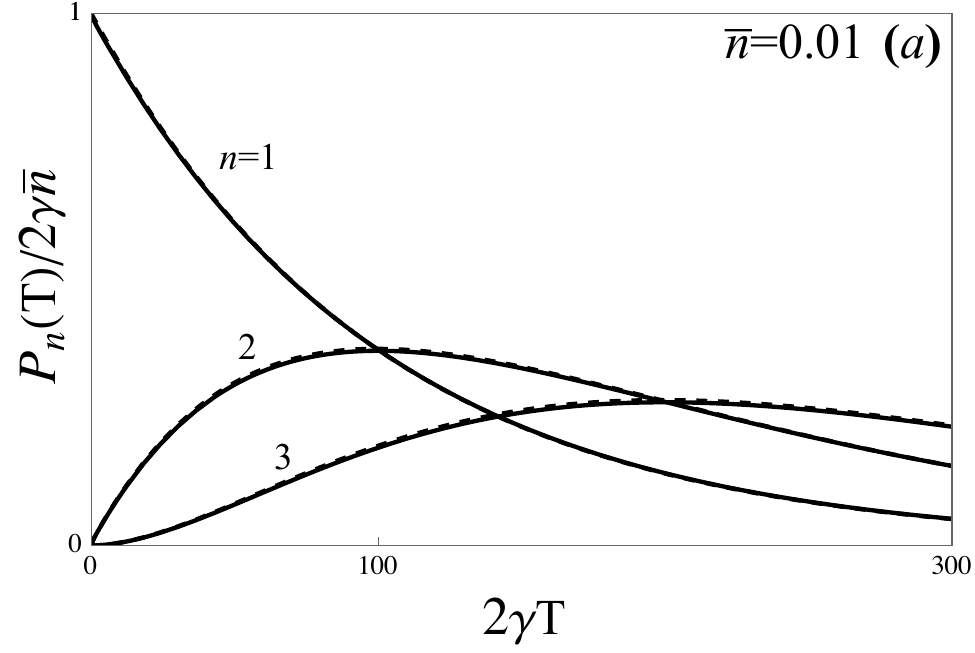}
 \includegraphics[width=0.45\linewidth]{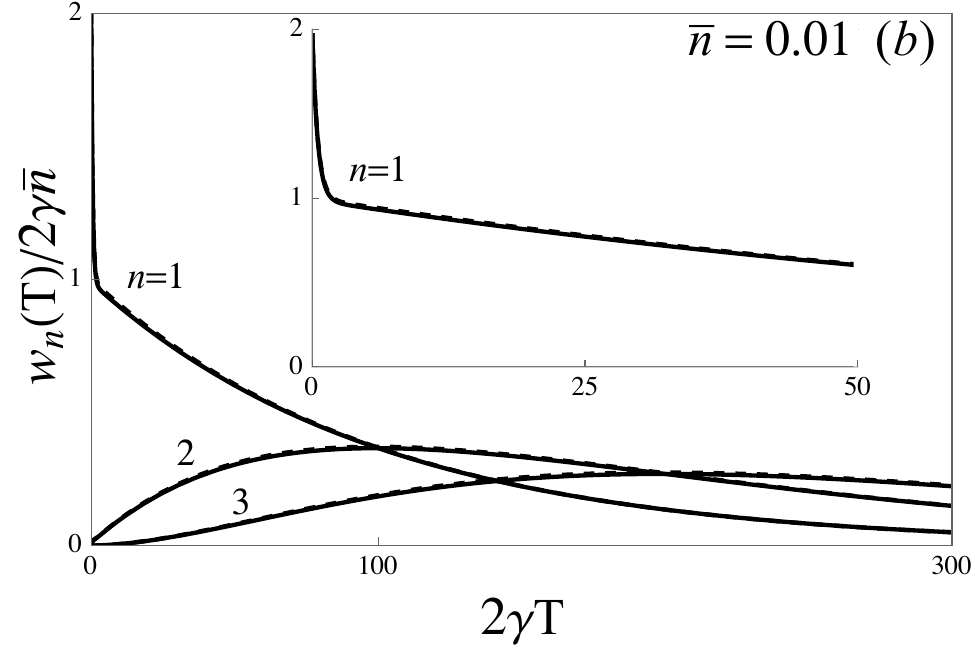}
    \caption{Comparison of the exact (solid curves) and approximate (dashed curves) expressions \eqref{P1th0}--\eqref{P3th0} for  $P_n(T)$ and \eqref{w1th0}--\eqref{w3th0} for $w_{n}(T)$ with $n=1,2,3$ for thermal light for small mean photon number $\overline{n} =0.01$. }
    \label{fig:1}
\end{figure*}

\begin{figure*}[t]
  \centering
    \includegraphics[width=0.45\linewidth]{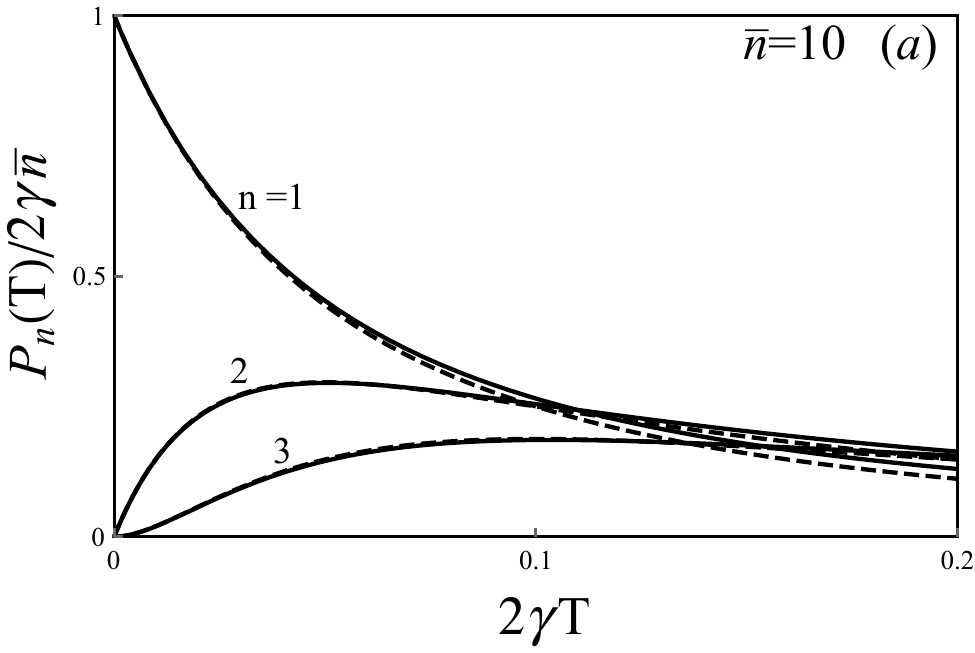}
    \includegraphics[width=0.45\linewidth]{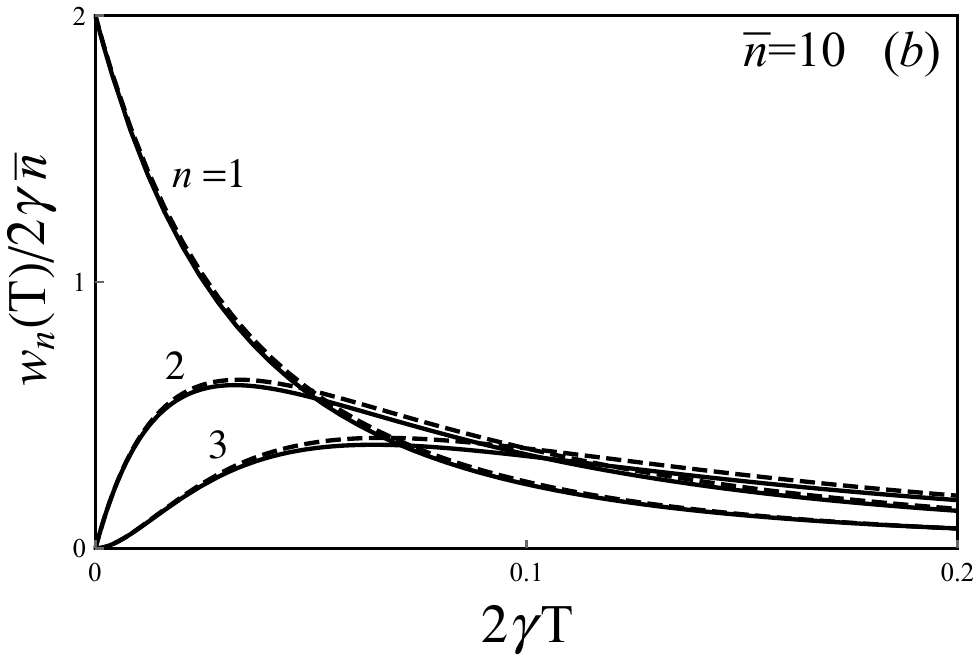}
    \caption{Comparison of the exact (solid curves) and approximate (dashed curves) expression \eqref{Pnth} for  $P_n(T)$ and \eqref{wnth} for $w_{n}(T)$ with $n=1,2,3$ for thermal light for large mean photon number $\overline{n} =10$.}
    \label{fig:2}
\end{figure*}

 \subsection{Large mean photon number $\bar{n}$}  A large mean photon number results in high photon flux from the source.  Therefore, the wait-time distributions are expected to be dominated by short wait-times. The generating function \eqref{Gth} for thermal light in this limit can be approximated by  \cite{saleh,vyassingh88}
\begin{equation}
G(s,T)  \approx \frac{1}{(1+s 2\gamma\eta \overline{n}  T)}\,.\label{gstth}
\end{equation}
Use of this generating function in Eqs.~\eqref{pnt} - \eqref{wnt2}, leads to the photo-count and wait-time distributions  
\begin{align}
p(n,T) &\approx \frac{( 2\gamma\eta \overline{n}  T)^{n}}{(1+\ 2\eta\gamma \overline{n}  T)^{n+1}}\,,\label{pth}\\
P_{n}(T) &\approx  {2\eta\gamma \overline{n}   } \frac{n\,(2\gamma\eta \overline{n}  T)^{n-1}}{(1+ 2\gamma\eta \overline{n}  T)^{n+1}}\,,\label{Pnth}\\
w_{n}(T) &\approx  {2\eta\gamma \overline{n}   } \frac{n(n+1)\,(2\gamma\eta \overline{n}  T)^{n-1}}{(1+ 2\gamma\eta \overline{n}  T)^{n+2}}\,.\label{wnth}
\end{align}
These approximations for the wait-time distributions satisfy the short time constraints \eqref{wnT=0}-\eqref{PnT=0}. They are compared with the exact expressions based on Eqs. (5) and (6) in Fig.~2, which shows excellent agreement for short times. 

For long wait-times $2\gamma T\gg1$, the integrated intensity $U(T)=\int_0^T {I}(t) dt \approx \langle I\rangle T=2\gamma\bar{n} T$,  so that the generating function, Eq.~\eqref{Gth}, takes the form 
\begin{equation}
G(s,T) 
 \approx e^{-2s\eta\gamma  \overline{n}  T} -\mathcal{O}(\overline{n}s)^{2}\,.
\end{equation}
This has the same form as the generating function \eqref{gstcoh} for coherent light. Thus, the long wait-time statistics are those of coherent light with mean photon flux $2\gamma\bar{n}$. Note that the wait-time distributions (16)--(21)  for small mean cavity photon number $\bar{n}$ have the expected exponential tail $e^{-2\eta\gamma\bar{n} T}$. The approximate wait-time distributions \eqref{Pnth}--\eqref{wnth} for large $\bar{n}$ do not fall off exponentially for large wait-times but they do capture the most significant part of the distributions and become increasingly accurate as $\bar{n}$ increases. This is the high degeneracy (large mode occupation number) limit of wait-time distributions \cite{saleh}.  

From Eqs. \eqref{pth}--\eqref{wnth},  we see that for large $\bar{n}$, the conditional and unconditional wait-time distributions for thermal light are related by 
\begin{align}\label{wnpnth} 
w_{n}(T) &\approx \frac{(n+1)}{(1+2\gamma \eta \overline{n} T)} P_{n}(T)\,.
\end{align}
We also note that both $w_1$ and $P_2$ are proportional to $g^{(2)}(0)$ for short times [Eqs. \eqref{wnT=0} and \eqref{PnT=0}], as both involve the detection of a pair of photons, but their behavior is quite different. The conditional wait-time distribution $w_1(T)$ has a maximum at $T=0$. This means, given that the counting begins at the detection of a photon, the first photo-detection, is most likely to occur immediately after the counting begins. In other words, thermal photons are bunched in time --  the detection of a photon makes the detection of the next photon most probable immediately after the first. In contrast, the unconditional wait-time distribution $P_2(T)$ vanishes at $T=0$. This means that if the counting begins at an arbitrary instant, the second photo-detection can occur only after a finite wait. It is clear there must be one photo-detection already before the second photo-detection can occur.  

The mean and variance of unconditional wait-time distributions can be evaluated  analytically in terms of hypergeometric functions \cite{luis23}. However, they will not be reproduced here. Instead, certain trends will be noted.  In the small $\bar{n}$ regime, the leading terms in the mean and variance have the same values as  coherent light of the same mean intensity. As $\bar{n}$ increases,  the mean and variance of wait-time decrease monotonically for all values of $n$.  A comparison of the most probable wait-times for thermal and coherent lights provides further insight into photon bunching in a thermal light beam. For coherent light, the most probable wait-time for the $n$th detection is $T_{n,\text{COH}} ={(n-1)}/{\eta\langle{I} \rangle}$. In contrast, for thermal light, the most probable wait-time, $T^{(P)}_{\text{n,\tiny TH}} = {(n-1)}/{2\eta \langle I \rangle}$, for unconditioned $n$th detection is shorter, and shorter still, $T^{(w)}_{n,{\text{\tiny TH}} }= {(n-1)}/{3\eta \langle I \rangle}$ for conditioned detection. This difference is to be expected due to bunching of photons in a thermal light beam compared to a random distribution of photons in time in a coherent beam. In particular, the wait-time $T_{1,\text{\tiny TH}}^{(w)}$ for conditioned detection of a photon pair in a thermal light beam is shorter than the wait-time $T_{2,\text{\tiny TH}} ^{(P)}$ for unconditioned detection of a photon pair in a coherent beam. 

Finally, we note that the effect of detection efficiency on wait-time distributions for thermal light is already included in the expressions given by Eqs. \eqref{P1th0}--\eqref{wnth}. Detection efficiency appears multiplied by $\bar{n}$ in the generating function. Thus, it's effect is to reduce photo-detection  rate from $2\gamma\bar{n}$ (for $\eta=1$) to $\eta 2\gamma\bar{n}$, without affecting the shape of the wait-time distributions.  
\section{Degenerate Parametric Oscillator}
The degenerate parametric oscillator (DPO) is perhaps the most important source of squeezed light \cite{kimble87}.  The basic mechanism of the DPO is the conversion of a pump photon into a pair of photons in a sub-harmonic mode (down-conversion) in a nonlinear medium inside an optical cavity.  Photons escaping from the cavity constitute the photon flux from the cavity. 

The light from the DPO requires a fully quantum mechanical treatment to describe its statistical properties.  Using the phase space representation of the field density matrix,  photon  annihilation and creation operators can be mapped onto two real independent Gaussian stochastic processes of zero mean but different variances \cite{drummond,wolinsky}. The generating function $G(s,T)$ for the light from a  DPO operating below threshold is then found to have the form  $ G(s,T)= Q_{1}(s,T) Q_{2}(s,T)$,
with $Q_{i}(s,T)$ given by \cite{vyassingh89-1,vyassingh89-2}:
\begin{align}
& Q_{i}(s,T) = \frac{e^{\lambda_{i} T/2}}{ \sqrt{\cosh(z_{i}T)+\frac{1}{2}\bigg[  \frac{z_{i}}{\lambda_{i}} + \frac{\lambda_{i}}{z_{i}} \bigg]\sinh(z_{i}T)}}, \label{Gdpo}
\end{align}
where $\lambda_{i} = \gamma \mp |\kappa \epsilon|,$, $z_{i}^{2} = \lambda_{i}^{2}\pm 2s\eta \gamma \kappa \epsilon$ for $i=\{1,2\}$, 
 $(2\gamma)^{-1}$ is the lifetime of the cavity, $|\epsilon|^{2}$ is proportional to the pump intensity, $s$ is a dimensionless parameter, and $\kappa$ is a coupling constant which depends on the properties of the nonlinear interaction between the pump and the down-converted photons inside the optical cavity. The mean photon number for the DPO cavity is given by $\bar{n} =(1/2)(|\kappa \epsilon/\gamma|^2)/(1-|\kappa \epsilon/\gamma|^2)$. This allows us to express the generating function  $G(s,T)$, more concisely, in terms of only $\gamma$ and $\bar{n}$.  It is interesting to note that thermal light is also described in terms of two real (classical) Gaussian processes with zero mean but both processes with the {\sl same} variance or equivalently, a single complex Gaussian stochastic process \cite{haken70}. This difference leads to very different quantum statistical properties for thermal light and the light from the DPO.

We first consider the unit detection efficiency($\eta =1$). The photo-detection sequence then is a direct representative of the photon sequence.  The general expressions for $P_{n}(T)$ and $w_{n}(T)$ cannot be written in terms of elementary functions and are not very illuminating \cite{luis23}. However, in several practically important limits, both $P_{n}(T)$ and $w_{n}(T)$  can be written in terms of simple analytic functions, which are excellent approximation to the exact expressions. These limits are discussed below. They allow us to construct a physical picture of photon sequences generated by this quantum mechanical source. 
\subsection{Small mean cavity photon number $\overline{n}$ } This corresponds to a small mean photon flux $2\gamma\bar{n}$ from the DPO cavity. The wait-times in this limit will be dominated by large intervals  $T \gg (2\gamma )^{-1}$. The generating function and the photo-count distribution in this limit take the form  \cite{vyassingh89-1,vyassingh89-2,huang}
\begin{align}
G(s,T)&\approx\left[1-\frac{\overline{n}}2s^2\right]\exp\left[\gamma\overline{n}T(s^2-2s)\right]\,,\label{gstdpo0}\\
p(2k,T)
&\approx  \frac{( \overline{n}\gamma T)^{k}}{k!} e^{-\overline{n}\gamma T}\,,\label{eq:p2ka}\\
p(2k+1,T) & \approx \overline{n}\frac{( \overline{n}\gamma T)^{k}}{k!} e^{-\overline{n}\gamma T}.\label{eq:p2k+1a}
\end{align} 
An inspection of Eqs. \eqref{eq:p2ka}  and \eqref{eq:p2k+1a} reveals a peculiar nature of photo-count distribution for the DPO: the probability of detecting an odd number of photo-counts is negligible compared to the probability of detecting an even  number of photo-counts ($p_{2k+1}/p_{2k}\propto\overline{n}\ll1$). It is as if the DPO emits photon pairs. This interpretation is reflected in the mathematical structure of Eqs. \eqref{eq:p2ka}  and \eqref{eq:p2k+1a}.  A cavity with mean photon number $\bar{n}$ will generate a photon flux $2\gamma\overline{n}$, resulting in $2\gamma\overline{n}T$ photo-counts in time interval $T$. This amounts to half as many photon-pairs $\overline{n}\gamma T$ in time  $T$.  A comparison of Eq.  \eqref{eq:p2ka}  with the corresponding expression \eqref{pncoh} for coherent light then shows that the probability of recording $2k$ photo-counts in DPO light  is the same as the probability of recording $k$ (random) photon-pairs. To appreciate this result, recall that the down-conversion of a pump photon results in a photon pair deposited inside the  DPO cavity, where each photon of a pair circulates independently, escaping the cavity in a lifetime of order $(2\gamma)^{-1}$. It is clear that the  creation of photon pairs inside the cavity alone is not sufficient  to give rise to a pair-like photo-count distribution of Eqs. \eqref{eq:p2ka}  and \eqref{eq:p2k+1a}. However, if the rate of photon-pair creation is sufficiently low,  both photons of a pair will escape the cavity before another pair is created.  A detector monitoring the output will then record, with high probability, an even number of photo-counts in a time $T$ large compared with the cavity lifetime $(2 \gamma)^{-1}$. 

 \begin{figure*}[t!]
  \centering
    \includegraphics[width=0.45\linewidth]{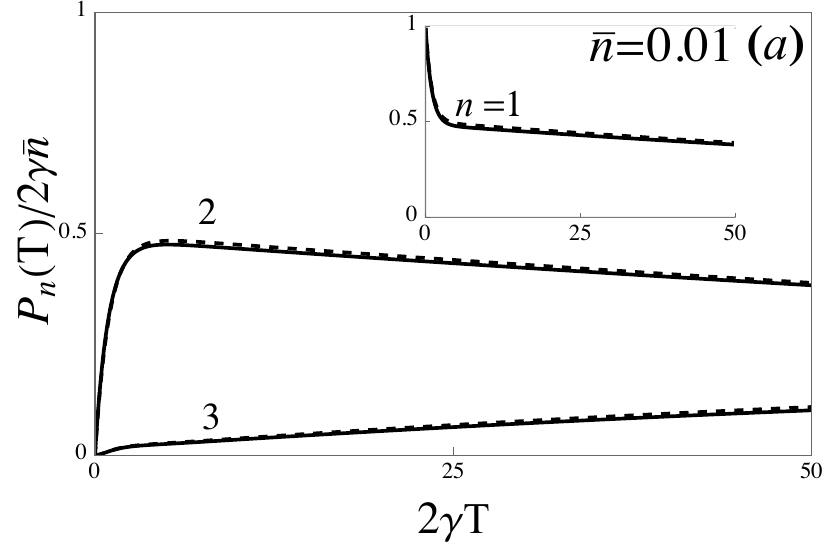}
     \includegraphics[width=0.45\linewidth]{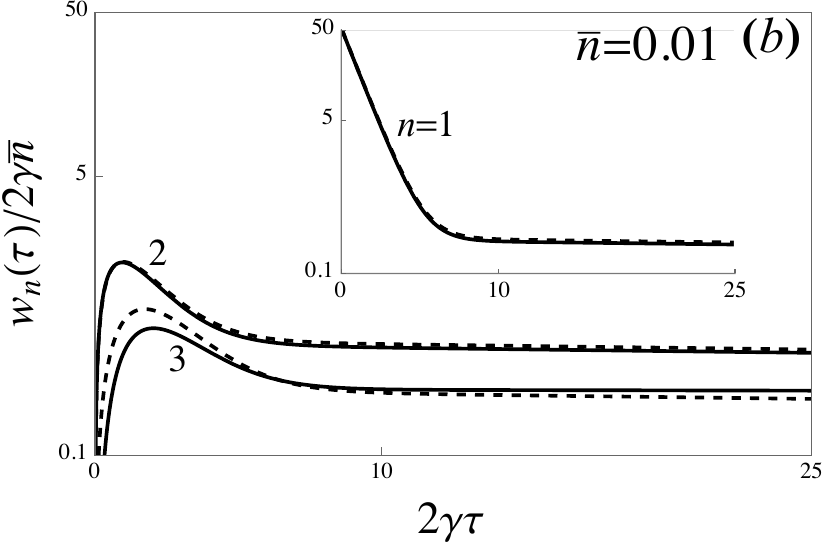}
    \caption{Exact wait-time distributions $P_{n}(T)$ and $w_{n}(T)$ for the light from the DPO for small cavity photon number $\overline{n}=0.01$.  The dashed curves are  derived from the approximate expressions in Eqs. \eqref{p1tdpo0}-\eqref{w3dpo0}.}
   \label{fig:3}
\end{figure*}

\begin{figure*}[t!]
  \centering
     \includegraphics[width=0.45\linewidth]{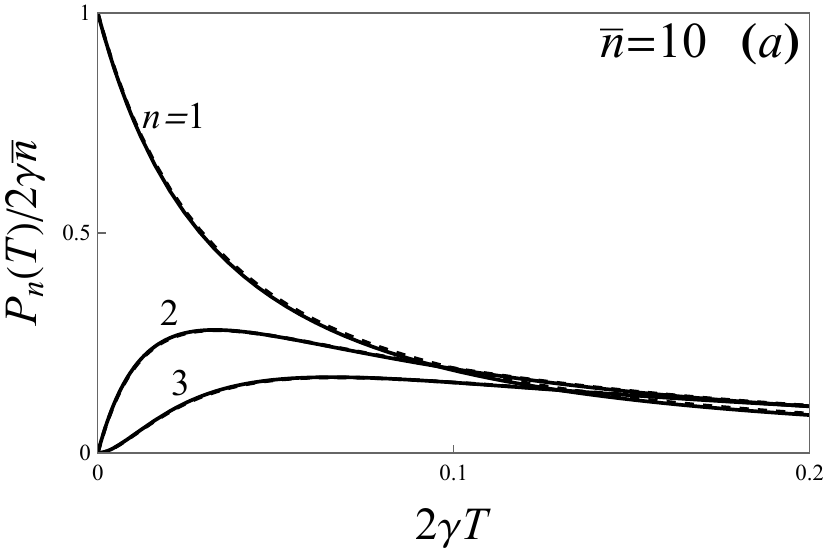}
    \includegraphics[width=0.45\linewidth]{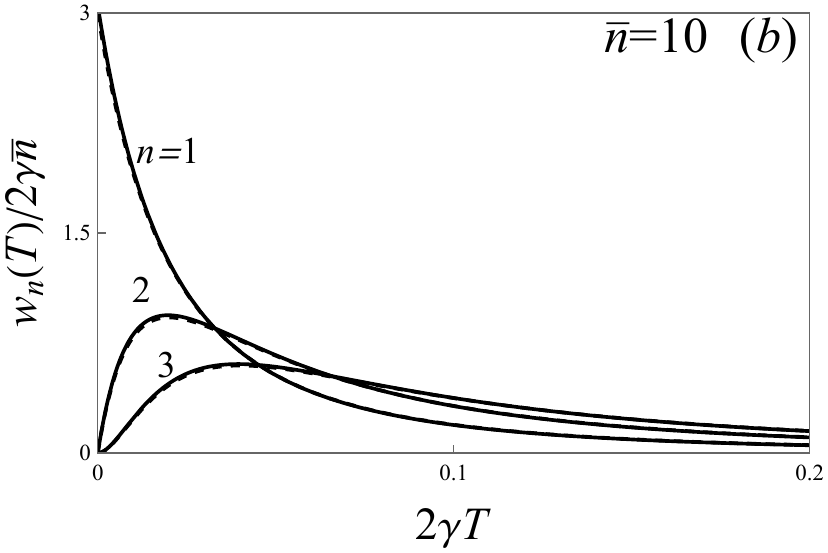}
    \caption{Exact curves for $P_{n}(T)$ and $w_{n}(T)$ for the light from the DPO  compared with approximate expressions (dashed) given by Eqs. \eqref{Pnbignbar} and \eqref{wnbignbar} for large mean  cavity photon number $\overline{n}=10$.}
   \label{fig:4}
\end{figure*}

This pair-like character of photoemission from the DPO  is reflected in the wait-time distributions as well. Equation \eqref{gstdpo0} correctly captures the long-time behavior of wait-time distributions but short time behavior, constrained by Eqs. (8) and (9) requires inclusion of more terms in $\bar{n}$ in the expansion of the generating function. As in the case of thermal light, this can be done relatively easily for $n=1-3$, which are sufficient for experimental characterization of photon sequences. The results for the unconditional distribution $P_n$ are 
\begin{align}
P_1(T)&\approx 2\gamma\bar{n}e^{-\gamma\bar{n}T}\frac12\left[1+e^{-2\gamma T}\right]\,,\label{p1tdpo0}\\
P_2(T)&\approx 2\gamma\bar{n}e^{-\gamma\bar{n}T}\frac12\left[1-e^{-2\gamma T}\right]\,,\label{p2tdpo0}\\
P_3(T)&\approx 2\gamma\bar{n}  e^{-\gamma\bar{n}T}\left(\frac{\bar{n}}2\right)\big[3+\gamma T+3e^{-4\gamma T}\notag\\
&\qquad\qquad  e^{-2 \gamma T} \left(6+\gamma T-4(\gamma T)^2  \right)  \big]\,.\label{p3tdpo0}
\end{align}
A similar procedure for the conditional wait-time distribution leads to 
\begin{align}
w_1(T)&\approx2 \gamma\bar{n}e^{- \gamma\bar{n}T} \left[ \frac{1}{4} +e^{-4\gamma T}+ e^{-2 \gamma T}\left(\frac{1}{2\bar{n}}+\frac{7}{4}\right)  \right],\label{w1dpo0}\\
w_2(T)&\approx 2\gamma\bar{n} e^{-\gamma \bar{n}T }\left[\frac12 -2e^{-4\gamma T}+ e^{-2\gamma T} \left( \frac32 +4\gamma T\right)\right],\label{w2dpo0}\\
w_3(T)&\approx 2\gamma\bar{n} e^{-\gamma\bar{n}T}\left[\frac14+e^{-4\gamma  T}\right.\notag\\
&\qquad\left.\qquad +e^{-2 \gamma T} \left( 2 (\gamma T)^2+\frac32 \gamma T-\frac54\right) \right]\,.\label{w3dpo0}
\end{align}
Equations  \eqref{p1tdpo0} - \eqref{w3dpo0} for the wait-time distributions are compared with the exact  distributions  in Fig. $3$. It can be seen that these approximate expressions capture both the short and long-time behavior as well as the general shape of the distributions. 

Several features of small $\bar{n}$ distributions given by Eqs. \eqref{p1tdpo0}-\eqref{w3dpo0} are noteworthy. First, the most significant qualitative difference between unconditional and conditional wait-time distributions appears in low order ($n=1-3$) distributions. For $n\ge 3$, the two types of wait-time  distributions have similar behavior in that they all vanish at $T=0$, reach a maximum at some nonzero value of $T$ and then decay exponentially to zero for large wait-times. Second, they are characterized by two very different time scales. The short time scale is the inverse of photo-emission rate $2\gamma$ (which  would correspond to photoemission rate with $\bar{n}=1$. This rate, which far exceeds the average photo-emission rate $2\gamma\bar{n}$,  can be thought of as photoemission rate conditioned upon a photo-detection. In this small $\bar{n}$ regime, the cavity has either a pair of photons (for a period lasting a few cavity lifetimes) or no photons. The detection of a photon in this regime signals, with high probability, the presence of one photon in the cavity, resulting in a photo-emission rate $2\gamma$ following a photo-detection. This is a manifestation of strong nonclassical correlations between the photons of a pair produced in the process of downconversion.  The long-time scale $ \sim(\gamma\bar{n})^{-1}$ is the inverse of mean photon pair emission rate. Indeed, the long wait-time tail of these distributions is Poissonian with mean flux  $\gamma\bar{n}$, which is half the mean photon flux $2\gamma\bar{n}$ from the cavity.  Also, for large wait-times, $2\gamma T\gg1$,  $P_1\approx P_2$ and $P_3\approx P_4$. This pattern extends to higher order distributions with $P_{2k+1}\approx P_{2k+2}$. Thus, the long wait-time behavior of wait-time distributions mimics a random photon sequence of mean flux $\gamma\bar{n}$ (not $2\gamma\bar{n}$).  Both of these aspects reflect the pair-like character of photoemissions from the DPO in the small $\bar{n}$ regime.

Another interesting feature is that the short time distributions for the DPO are ``super-thermal." To appreciate this, recall that for short times,  the wait-time distributions are proportional to zero-delay intensity correlation functions  [Eqs. \eqref{wnT=0}--\eqref{PnT=0}]. For example, $w_1(T)\propto \langle:{\hat I}^2:\rangle $, which for thermal light is $\sim\bar{n}^2$  \cite{saleh}, whereas for the light from the DPO it is $\sim\bar{n}$ \cite{friberg85,vyassingh89-1}. Since $\bar{n}\gg\bar{n}^2$ for $\bar{n}<1$, it follows that $g^{(2)}_{_\text{\tiny DPO}}(0)>g^{(2)}_{_\text{\tiny TH}}(0)$. Similar results hold for all other distributions except $P_1(T)$, which, by definition, is normalized to $P_1(0)/ 2\gamma\bar{n}=1$. This can also be seen by comparing  Fig. 1 for thermal light and  Fig. 3 for the light from the DPO. 

\begin{figure*}[t!]
  \centering 
    \includegraphics[width=0.45\linewidth]{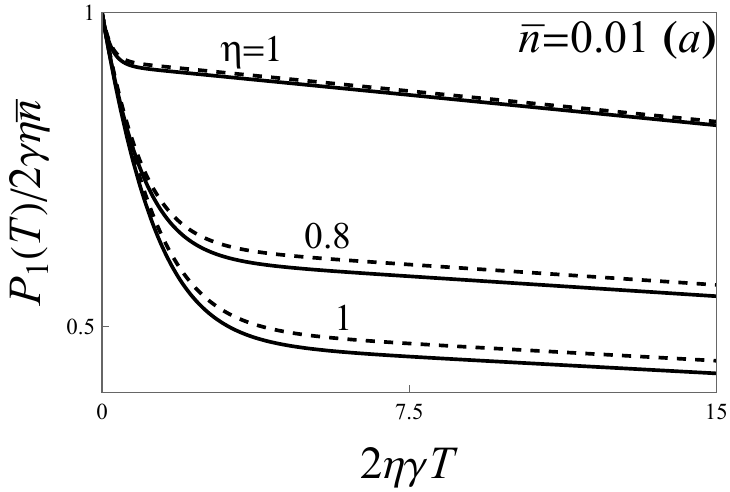}
    \includegraphics[width=0.46\linewidth]{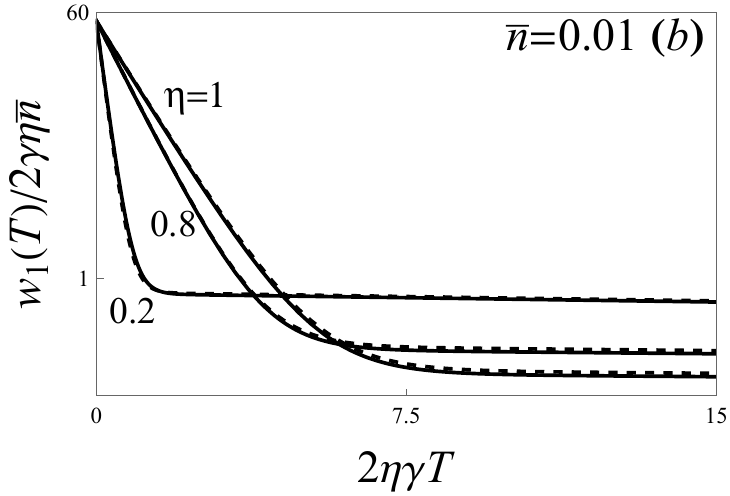}
    \caption{ Effect of detection efficiency on wait-time distributions for the DPO for small mean photon number ($\overline{n}=0.01$). The dashed curves show the approximations of Eqs. (48)-(51). }
   \label{fig:5}
\end{figure*}
\subsection{Large mean cavity photon number $\overline{n}$}  The large photon number regime  is also of practical interest. As the oscillator approaches threshold of oscillation, many photon pairs are created inside the cavity in a cavity lifetime and the photons escaping the DPO cavity cannot be interpreted as coming from the same pair \cite{vyassingh89-1}. In this case, the wait-times are dominated by  intervals small compared to the cavity lifetime ($2 \gamma )^{-1}$ and the generating function $G(s,T)$ can be approximated by \cite{vyassingh89-1,vyassingh89-2}
\begin{equation}
G(s,T)\approx \frac1{(1+4s\eta \gamma\bar{n}T)^{1/2}}\,,\label{gstdpo1}
\end{equation}
Using this in Eqs.~\eqref{Pnt2} and \eqref{wnt2} we find that the $P_{n}(T)$ and $w_{n}(T)$ can be obtained by differentiation,  leading to
\begin{align}
P_{n}(T)&\approx 2\eta\gamma \overline{n}\frac{(2n-1)!!}{(n-1)!} \frac{(2\eta\gamma\bar{n}T)^{n-1}}{(1+4\eta\gamma\bar{n}T)^{n+1/2}} \label{Pnbignbar}\,,\\
w_{n}(T)&\approx 2\eta\gamma \overline{n}\frac{(2n+1)!!}{(n-1)!}\frac{(2\eta\gamma\bar{n}T)^{n-1}}{(1+4\eta\gamma\bar{n}T)^{n+3/2}}, \label{wnbignbar}
 \end{align}
Equations~\eqref{Pnbignbar} and  \eqref{wnbignbar} are compared with the exact curves in Fig.~4 and can be seen to be good approximations to the exact distributions. A comparison of these curves with those in Fig. 3 for small $\bar{n}$, shows that the most significant change in the wait-time distributions with increased cavity photon number $\bar{n}$  is that the two very different time scales $(2\gamma)^{-1}$ and $(\gamma\bar{n})^{-1}$, the former corresponding to enhanced cavity emission following the detection of the first photon and the latter corresponding to separation between photon pairs, which were so prominent at small cavity photon numbers, have been replaced by a single time scale $(2\eta\gamma\bar{n})^{-1}$ determined by the mean cavity photoemission rate $2\gamma\bar{n}$. This is the so-called high degeneracy limit of squeezed light discussed in Ref. \cite{vyassingh88}.  

The wait-time distributions of Eqs. \eqref{Pnbignbar} and \eqref{wnbignbar} are related by 
\begin{align}\label{wnpndpo} 
w_{n}(T) &= \frac{(2n+1)}{(1+4 \eta\gamma\bar{n} T)}P_{n}(T)
\,, \end{align}
which is remarkably similar to  the relation \eqref{wnpnth} for thermal light. In fact, for large mean cavity photon number, the curves of Fig. 4 are qualitatively similar to those in Fig. 2 for thermal light. However, a closer examination of Eqs. (27) and (41) and the curves in Figs. 2 and 4 reveals quantitative differences. The conditional wait times for the DPO are biased toward shorter times compared to thermal light - they are peaked at shorter wait times and are narrower than those for thermal light. For example, $w_1(0)$ for the DPO is $3/2$ times as large as that for thermal light.  Indeed, the most probable conditioned wait-time for the $n$th photo-detection, $T^{(w)}_{\text{\tiny DPO}}= (n-1)/5\langle \hat{I} \rangle$, for the DPO is shorter than the corresponding time for thermal  photons, $T^{(w)}_{_\text{\tiny TH}}= (n-1)/3\langle \hat{I} \rangle$, where $\langle \hat{I} \rangle$ is the mean photon flux.
\subsection{Effect of detection efficiency} 
Non-unit detection efficiency ($\eta<1$) causes the photo-detection sequence to differ from the photo-emission sequence. The most significant effect of the non-unit detection efficiency for the DPO is to degrade the even-odd oscillations in the photo-count distribution. This has been discussed in detail in Ref. \cite{vyassingh89-1}. For non-unit detection efficiency, the expressions for $P_{n}(T)$ and $w_{n}(T)$ for arbitrary $n$ do not have simple forms, in general. However, for large  $\bar{n}$, the dominant effect of detection efficiency is already contained in expressions \eqref{Pnbignbar} and  \eqref{wnbignbar}  and is similar to that found for thermal light in the high degeneracy limit \cite{vyassingh88}. 

For small $\bar{n}$, where quantum effects dominate, the effect of detector efficiency is more interesting and illustrates how the quantum nature of photo-emission sequence can be obscured in the photo-detection sequence. As noted earlier, wait-time distributions beyond $n=1$ and 2 carry little qualitatively new information. Therefore, we will limit our considerations of non-unit detection efficiency to $n=1,2$ distributions. Following a procedure similar to that used in arriving at Eqs. \eqref{p1tdpo0}-\eqref{w3dpo0} for small $\bar{n}$, we expand the generating function in powers of $\bar{n}$ and retain terms necessary to satisfy the constraints of Eqs. \eqref{wnT=0}-\eqref{PnT=0}.   We then obtain the following expressions
\begin{align}
P_1(T)&\approx2\eta\gamma\bar{n}e^{-(2\eta-\eta^2)\gamma\bar{n}T}\frac12\left[ 2-\eta+\eta e^{-2\gamma T}\right]\,,\label{p1tdpononunit}\\
P_2(T)&\approx2\eta\gamma\bar{n}e^{-(2\eta-\eta^2)\gamma\bar{n}T}\frac{\eta}2\left[ 1-e^{-2\gamma T}\right]\,,\label{p2tdpononunit}\\
w_1(T)&\approx2\eta\gamma\bar{n}e^{-(2\eta-\eta^2)\gamma\bar{n}T} \left[\frac14(\eta-2)^2+\eta^2e^{-4\gamma T}\right.\notag\\
&\qquad\left. +e^{-2\gamma T}\left(2+\frac1{2\bar{n}}+\eta -\frac54\eta^2\right)\right]\label{w1tdpononunit}\\
w_2(T)&\approx2\eta\gamma\bar{n}e^{-(2\eta-\eta^2)\gamma\bar{n}T}\frac{\eta}2 \left[ 2-\eta-4\eta e^{-4\gamma T}\right.\notag\\
&\quad\left.+e^{-2\gamma T}\left(5\eta-6\eta\gamma T+14\gamma T-2\right)\right].\label{w2tdpononunit}
\end{align}
As a check, we note that for $\eta=1$, these expressions  reduce to those in Eqs. \eqref{p1tdpo0} -\eqref{w3dpo0} and satisfy the constraints \eqref{wnT=0}-\eqref{PnT=0}. Noteworthy is the nonlinear dependence of wait-time distributions, especially, the exponents, $(2\eta-\eta^2)\gamma \bar{n} T$, on detection efficiency. Thus, the overall effect of detection efficiency on wait-time distributions goes beyond simple scaling of wait-time distributions.

Figure 5 illustrates the effect of non-unit detection efficiency on wait-time distributions. The full curves represent numerical calculations using the generating function (28) in Eqs. (5) and (6). The dashed curves are obtained from Eqs. \eqref{p1tdpononunit} - \eqref{w2tdpononunit}.  Note that as $\eta$ decreases, the long-time scale determined  by the pair-emission rate $\gamma\bar{n}$ for $\eta =1$ is replaced by a time scale determined by the mean photon flux $2\gamma\bar{n}$. We can also see analytically, from the exponential before the square brackets, that for small detection efficiency, the exponent $(2\eta-\eta^2)\gamma\bar{n} T\to2\eta\gamma\bar{n}$. In this quantum regime, the detection efficiency changes the time scales as well as the shape of the distributions. As the detection efficiency decreases, the observed wait-time distribution resembles a rate limited distribution for classical (thermal and coherent) light. Here we have an analytical model that allows us to see how the quantum mechanical properties of a photo-emission sequence are washed out in the photo-detection sequence as detection efficiency decreases.   
\section{Two-Level Atom Resonance Fluorescence}
Consider now the photon sequence produced by a two-level atom driven by a coherent field with frequency close to the atomic resonance frequency. With each photoemission, the atom returns to its lower state, and so each subsequent photon-emission occurs with the atom starting in the lower state, independent of the history of previous photoemissions. The driving field, being in a classical (coherent) state, remains unaffected. This property allows the averages of products of photon flux operators to be simplified and expressed in terms of the products of two-time averages.  This fascinating example of a quantum mechanical light source has been studied in detail  \cite{carmichael76,kimble76,mandel77,singh83,carmichael93,lenstra,arnoldus} using photo-count distribution as well as the wait-time distributions \cite{carmichael89,arnolduswn,arnolduswn2}. We have generalized these to arbitrary $n$ and efficiency of detection. The details of this are relegated to the Appendix.   Here we simply quote the steady-state results for the conditional wait-time distributions $w_n(T)$ for $n=1,2,3$ to allow a comparison with light from the DPO.  
 
 \subsection{Wait-time distributions}
The  conditional wait-time distributions for $n=1-3$ in the steady-state are given by [Appendix and Ref. \cite{arnolduswn}]
 \begin{align}
&w_1(T)=  \frac{\varOmega^2}{\omega^2}e^{-\beta T} \bigg[-1+\cosh\omega T \bigg], \label{w1rf}\\
&w_{2}(T)= \frac12  \frac{\varOmega^4}{\omega^4}\beta T e^{-\beta T}\bigg[2+\cosh\omega T-3\frac{\sinh\omega T}{\omega T}\bigg], \label{w2rf}\\
&w_{3}(T)=\frac18  \frac{\varOmega^6}{\omega^6} (\beta T)^2 e^{-\beta T}
\bigg[-4+\cosh\omega T-9\frac{\sinh\omega T}{\omega T}\nonumber \\
&\qquad+24\frac{\cosh\omega T-1}{(\omega T)^{2}} \bigg], \label{w3rf}
\comment{ \\
&w_{4}(T)= \frac{C_{4}}{3 \cdot 2^{4}}( \beta T)^3 e^{-\beta T}
\bigg[8+\cosh\omega T-18\frac{\sinh\omega T}{\omega T}\nonumber \\
&\quad+\frac{3}{(\omega T)^{2}}\bigg(64+41\cosh\omega T -105\frac{\sinh\omega T}{\omega T} \bigg)\bigg], \label{w4rfweak}}
\end{align}
where $2 \beta$ is the Einstein-\textit{A} coefficient, $\varOmega$ is the Rabi frequency for the atomic transition, and $\omega = \sqrt{\beta^{2}-\varOmega^{2}}$. 
These expressions written for Rabi frequency $\varOmega<\beta$ also hold for $\varOmega>\beta$ if we replace the hyperbolic functions  by trigonometric functions: $\cosh(\omega T)\to \cos\, |\omega| T$ and $ \sinh(\omega T)/(\omega T) \to \sin( |\omega|T)/(|\omega| T)$. 

\begin{figure*}[t!]
  \centering
    \includegraphics[width=0.4\linewidth]{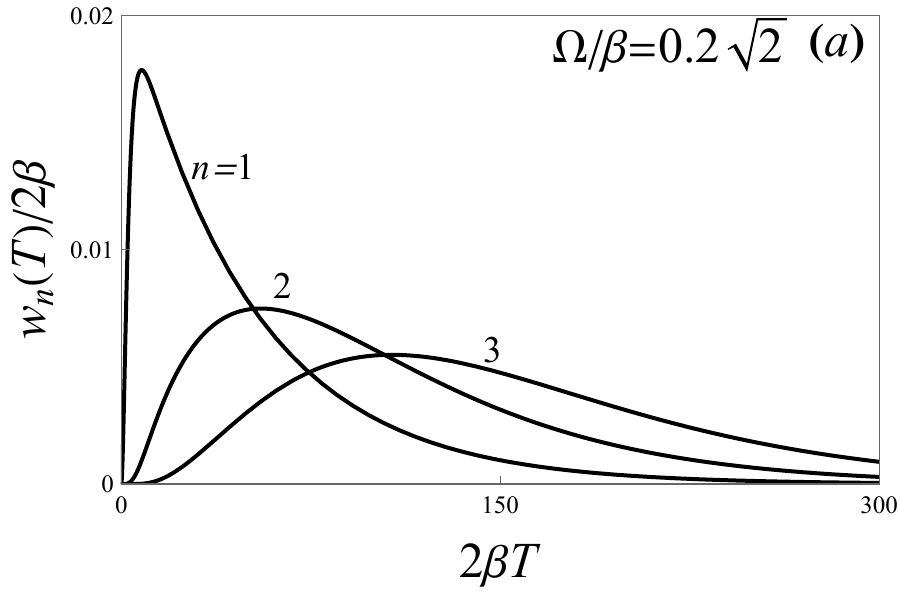}
        \includegraphics[width=0.4\linewidth]{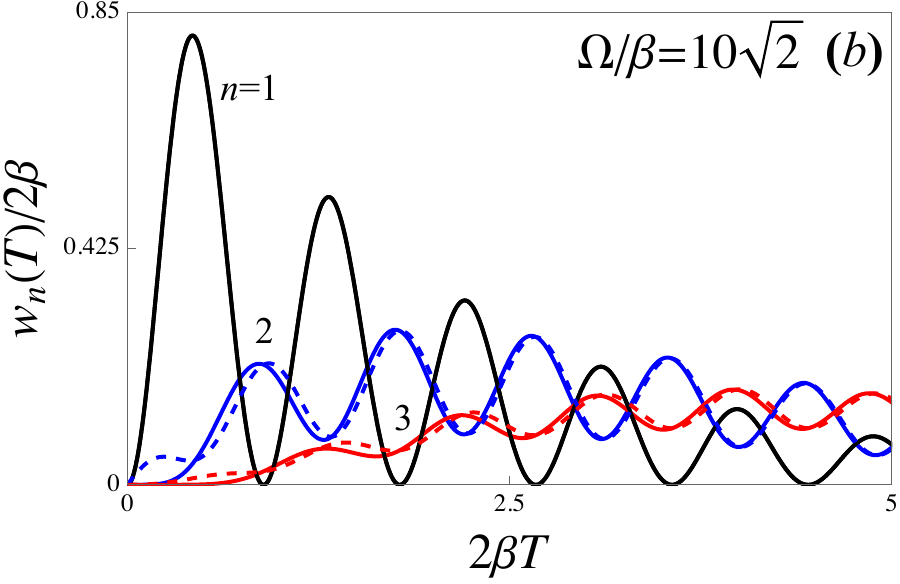}
      \includegraphics[width=0.4\linewidth]{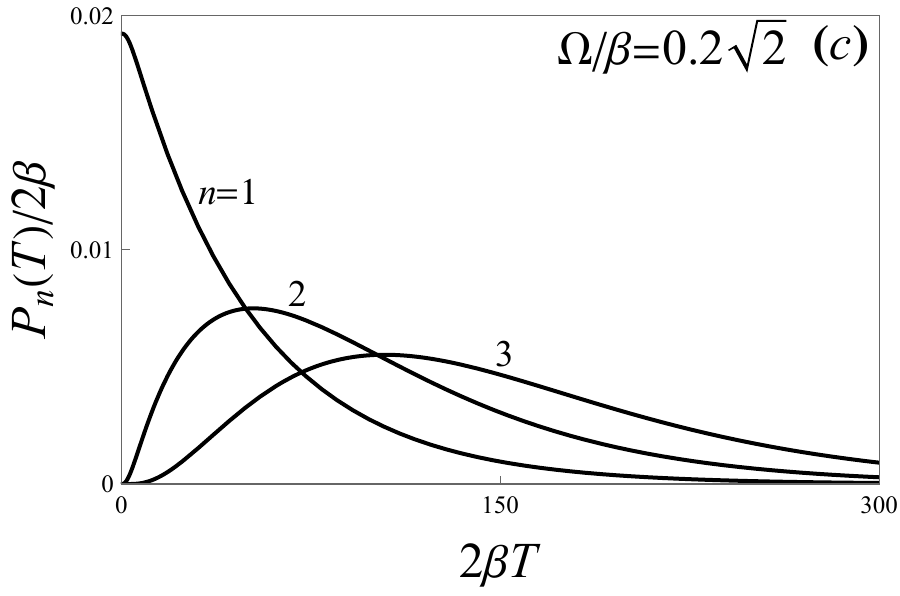}
        \includegraphics[width=0.4\linewidth]{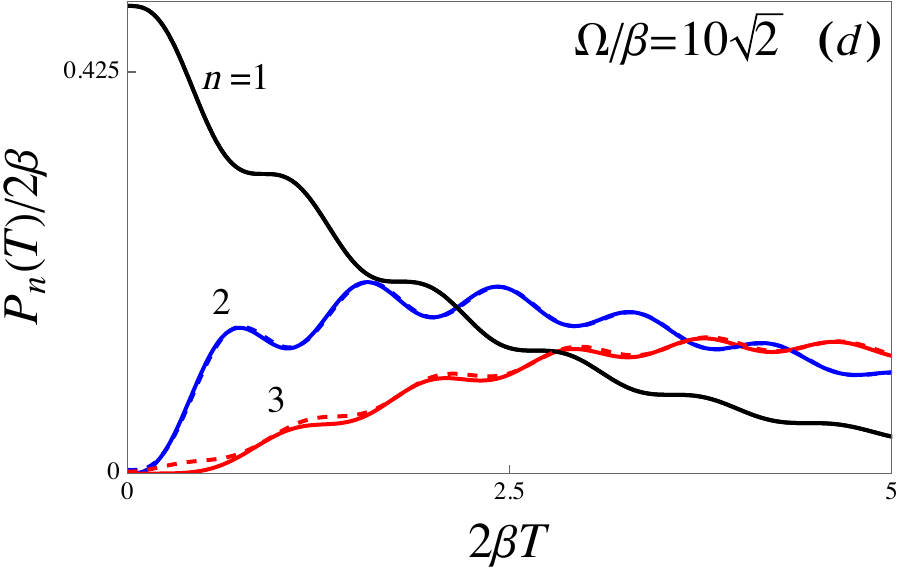}
    \caption{$w_n(T)$ and $P_n(T)$ for small ($\varOmega/\beta=0.2\sqrt{2}$) and large ($\varOmega/\beta=10\sqrt2$) driving fields for $n=1-3$. The dashed curves in frames (b) and (d) represent large field approximations [Eq. \eqref{wnrfstrongfield1}]. 
     }
   \label{fig:6}
\end{figure*}
These wait-time distributions are shown in Fig. 6  for two values of Rabi frequency  $\varOmega<\beta$ ($\varOmega/\beta=0.2\sqrt2$) and $\varOmega>\beta$ ($\varOmega/\beta=10\sqrt2$).  In all cases, the conditional wait-time distributions vanish at $T=0$ and reach a single maximum (or more) at some nonzero time before decreasing to zero as $T$ increases. The behavior of $w_n$ for short times, $2 \beta T \ll 1$,  can be obtained by a Taylor expansion of the terms inside  square brackets in Eqs. \eqref{w1rf} - \eqref{w3rf}, giving us the leading term 
\begin{align}
w_{n} \approx \frac{\Omega^{2n}}{\beta^{2n}} \beta \frac{(\beta T)^{3n-1}}{(3n-1)!}e^{-\beta T}. \label{wnrfsmallT}
\end{align}
Thus, for short wait-times, $w_n$ is proportional to a gamma distribution
 with shape parameter $3n$ and rate $\beta$  \cite{arfken}. This short-time behavior of $w_{n}(T)$ differs from that derived in Eq. \eqref{wnT=0} for the light from the DPO or the thermal light from a laser operating below threshold. This is because the zero-delay intensity correlation functions $g^{(n)}(0)$ ($n\ge2$)  vanish in resonance fluorescence. Therefore, the leading nonzero term in the short-time limit requires a calculation carried out to a higher order in the small parameter $\beta T$. Physically, the vanishing of $w_n(0)$ reflects the fact that a photo-detection signals the return of the atom to its ground state and, therefore, its inability to emit another  photon immediately after a photo-detection has occurred. 

The behavior of wait-time distribution $w_n$ changes qualitatively as the Rabi frequency $\varOmega$ increases.  For $\varOmega<\beta$, the distributions have a single maximum, whereas for $\varOmega>\beta$, they begin to develop modulations reflecting Rabi oscillations. $w_1(T)$ has the strongest modulations and vanishes periodically at intervals that are multiples of $2\pi/|\omega|=2\pi/\sqrt{\varOmega^2-\beta^2}$. These zeros signal the atom's periodic return to the ground state when it is unable to emit a photon. Higher order distributions $w_2(T)$ and $w_3(T)$ also exhibit modulations but of smaller amplitude and they do not vanish except at $T=0$.  

For a strong driving field, $ {\Omega^2}/{\beta^2} \gg1$,  the steady-state emission rate  
$\langle \hat{I} \rangle_{ss} = {\beta \Omega^{2}}/({\Omega^{2}+2\beta^{2}})\to \beta$ and the wait-time distributions $w_{n}(T)$ simplify to
\begin{equation}
w_{n}(T) \approx\beta  \frac{(\beta T)^{n-1}}{(n-1)!}{e^{-\beta T}}\bigg[1-\frac{(-1)^{n-1}}{2^{n-1}}\cos(\Omega T)\bigg]. \label{wnrfstrongfield1}
\end{equation}
This is a gamma distribution modulated at Rabi frequency, the modulations of consecutive distributions being $\pi$ out of phase. This approximation is shown by the dashed curve in Fig. 6(b), which appears to reproduce the exact curves for both $w_n$ beyond the first maximum quite well.  
 
The unconditional wait-time distribution $P_{n}(T)$ can be derived using a procedure similar to that used for $w_n(T)$ or by using Eq. \eqref{Pnrf1} of  Appendix.  The resulting expressions are similar to those for $w_{n}(T)$ and we omit writing their explicit form here. 
Figures~6(c) and 6(d) show $P_n(T)$ for small and large driving fields. Their comparison with conditional distributions shows that  $P_1(T)$ and  $w_1(T)$ differ the most from each other; the latter vanishes at $T=0$ whereas $P_1(T)$ has $T=0$ as the most probable value. It is interesting to note that although $P_1(T)$ corresponds to the first photo-detection when counting starts at a random instant, the underlying Rabi oscillations are not completely washed out [Fig. 6(d)].  For $n\ge2$, the unconditional and conditional distributions have qualitatively similar behavior.  Both types vanish at $T=0$ and reach a single maximum for $\varOmega/\beta <\sqrt2$ or exhibit modulations for $\varOmega/\beta>\sqrt2$ before decreasing to zero.    
\subsubsection*{A Special case}
A comparison of the wait-time distributions in resonance fluorescence with those for thermal and DPO light (Figs. 1- 4) shows that for $\varOmega/\beta<\sqrt2$ they share a common feature by having a single maximum. For a quantitative comparison, we consider the special case $\Omega/\beta = 1$ in this regime. The steady-state photon flux in this case is  $\langle I\rangle_{ss}=\beta\varOmega^2/(\varOmega^2+2\beta^2)\to \beta/3$ and  Eq.~\eqref{w1rf} - \eqref{w3rf} simplify to  
 \begin{equation}
w_{n}(T) = \beta \frac{(\beta T)^{3n-1}}{(3n-1)!}e^{-\beta T}\,.
 \label{wnRF1}
\end{equation}
The corresponding expression for the unconditional distribution is  
\begin{align}
P_{n}(T)&= \frac{\beta}{3} \frac{(\beta T)^{3(n-1)}}{(3n-1)!} e^{-\beta T} \times \nonumber \\& \big[(\beta T)^{2} +(3n-1)\beta T +(3n-1)(3n-2)\big].\label{pnrf1}
\end{align}
Distributions \eqref{wnRF1} and \eqref{pnrf1} are shown in Fig. 7 for $n=1-3$ as a function of $T$. They all have a single maximum, which occurs  at increasingly larger values of $T$ as $n$ increases. Equation  \eqref{wnRF1}  is a gamma distribution with shape parameter $3n$, rate $\beta$ and wait-time average $\langle T\rangle_{w_n}={3n}/{\beta}=n/\langle I\rangle_{ss}$ and variance $\langle(\Delta T)^2\rangle_{w_n}=3n/\beta^2=n/3\langle I\rangle_{ss}^2$ \cite{arfken}. The mean is the same as that for a Poisson sequence of the same flux $\langle I\rangle_{ss}$ but the variance is smaller than the variance for a Poisson sequence [Eqs. \eqref{tavcoh} and \eqref{tvarcoh}]. The sub-Poissonian variance of wait-times implies that the photoemissions in resonance fluorescence are more {\it regular} than a Poisson sequence. The regular nature of photo-emissions is reflected in the most probable wait-time $\tau_{_\text{RF}}=(3n-1)/\beta=(n-\frac13)\langle \hat{I} \rangle_{ss}^{-1}$ for  the $n$th photo-detection being longer than the average wait-time $\tau_{_\text{COH}} = (n-1)\langle \hat{I} \rangle_{ss}^{-1}$ for a Poisson sequence [Eq. \eqref{Pncoh}]. These conclusions, though reached by considering the special case $\varOmega/\beta=1$, hold generally. This is discussed further, for arbitrary values of detection efficiency and $\varOmega/\beta$ in the next subsection.  

\begin{figure*}[t!]
  \centering
    \includegraphics[width=0.4\linewidth]{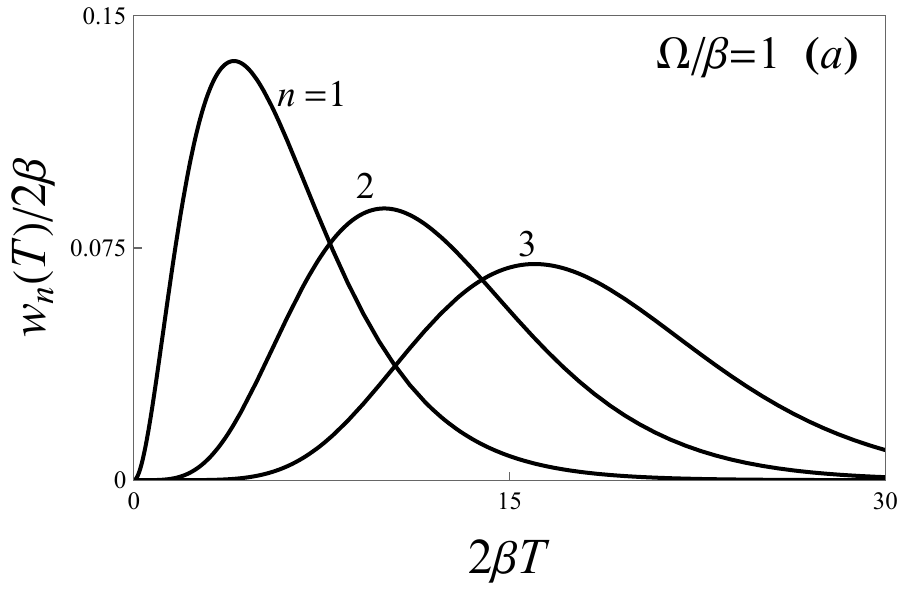}
    \includegraphics[width=0.4\linewidth]{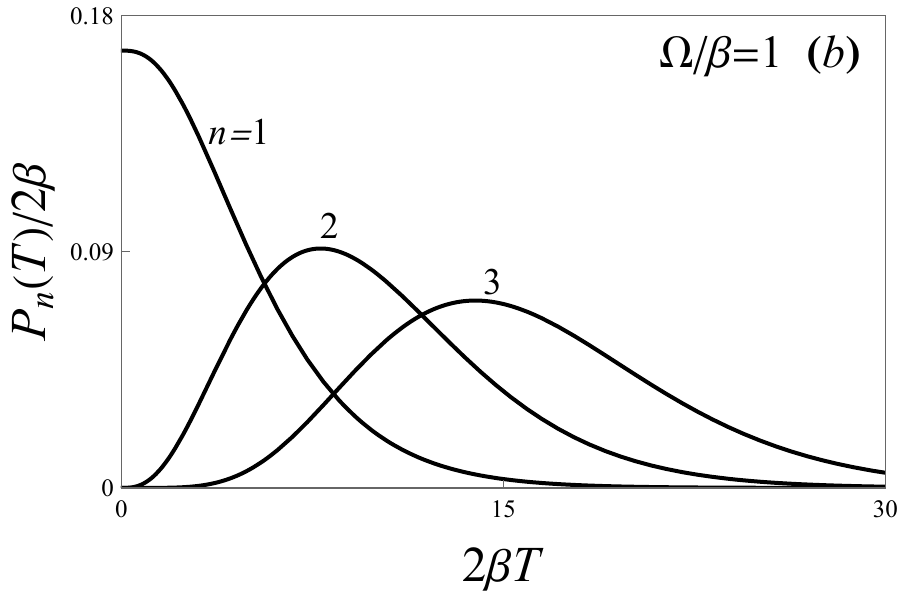}
 \caption{$w_{n}(T)$ and $P_{n}(T)$ for $n=1-3$ in resonance fluorescence  for $\Omega = \beta$. }
   \label{fig:7}
\end{figure*}

 \begin{figure*}[t!]
  \centering
    \includegraphics[width=0.4\linewidth]{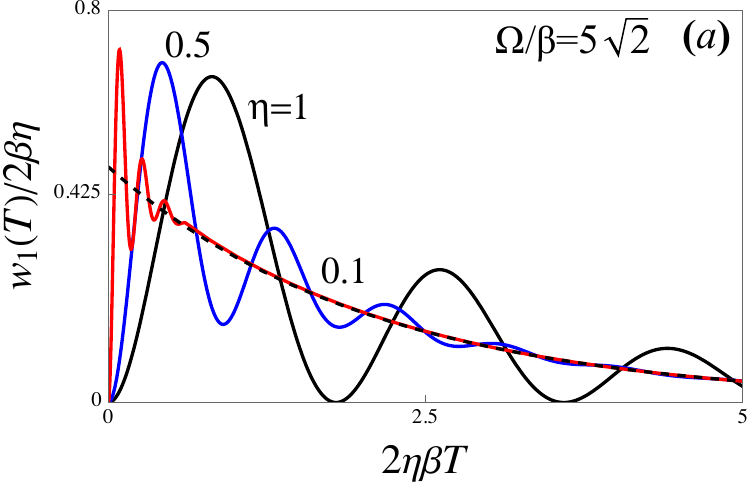}
    \includegraphics[width=0.4\linewidth]{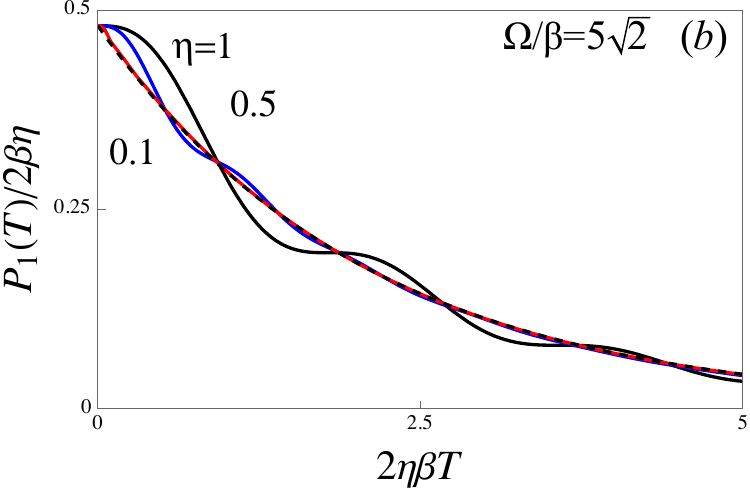}
    \includegraphics[width=0.4\linewidth]{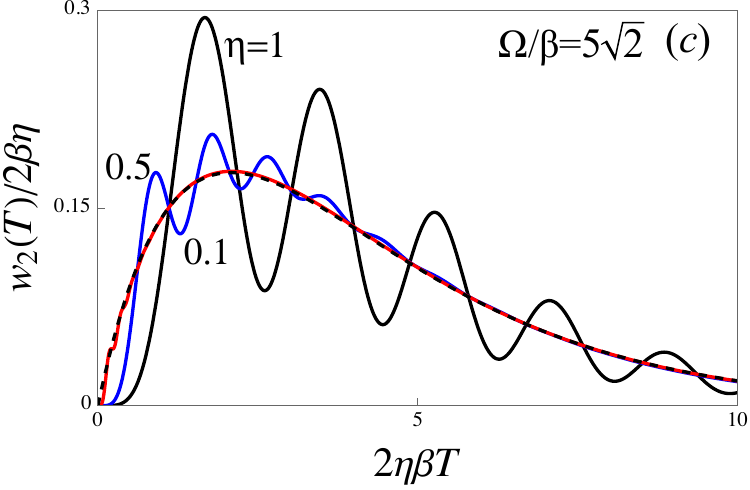}
    \includegraphics[width=0.4\linewidth]{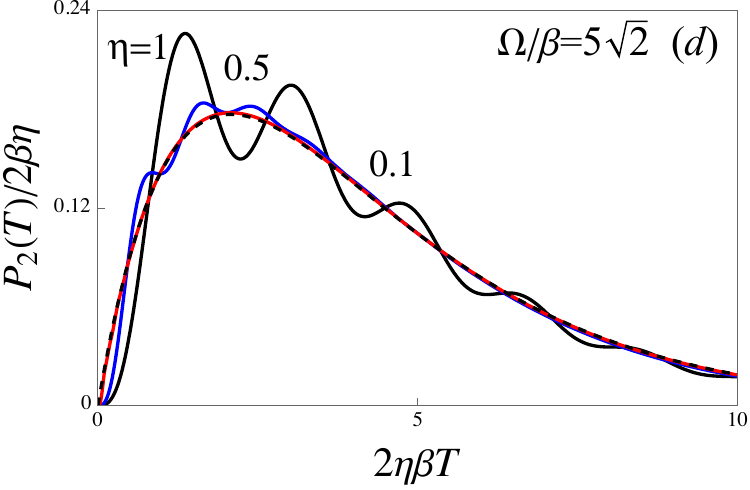}
    \caption{Effect of non-unit detection efficiency on wait-time distributions in resonance fluorescent for n = 1, 2 and efficiency $\eta=1$(black), 0.5 (blue) and 0.1 (red).  The full curves are derived from Eq. (84). The dashed curve in each frame represents the distribution for coherent light.}
   \label{fig:8}
\end{figure*}

\subsection{Effect of detection efficiency} 
For non-unit detection efficiency, the expressions for $w_n(T)$ and $P_n(T)$ become cumbersome.  The exact formulas have been relegated to  the Appendix. Here we illustrate the effect of non-unit detection efficiency on wait-time distributions by considering some special cases.  

For non-unit detection efficiency, the wait-time distribution $w_1(T)$ for $\Omega = \beta$ is given by 
  \begin{align}
w_{1}(T)= &\frac{\eta \beta}{3 \mu^{2}}e^{-\beta T(1-\mu)}  \bigg[1 \notag\\
&\quad -2e^{-\frac{3}{2}\beta \mu T}\cos \bigg( \frac{\sqrt{3}}{2} \mu \beta T-\frac{\pi}{3} \bigg)\bigg],\label{w1rfnonunit}
\end{align} 
where $\mu =(1-\eta)^{{1/3}}$. This reduces to $w_1(T)$ given by Eq. (51) for $\eta=1$ and vanishes for $T=0$, independent of the efficiency of detection, since $w_{1}(0)\propto  g^{(2)}(0)$.  On using Eq. (\ref{Pnrf1}) of the Appendix, we find that the unconditional wait-time distribution for the first photo-detection is 
\begin{align}
&P_{1}(T) = \frac{\eta \beta e^{-(1-\mu)\beta T}}{9 \mu^{2}}\bigg[(1+\mu+\mu^{2})\nonumber \\ 
&\quad -2e^{-\frac{3}{2}\mu \beta  T}\bigg\{ \bigg( 1-\frac{\mu}{2}-\frac{\mu^{2}}{2}\bigg)\cos \bigg( \frac{\sqrt{3}}{2} \mu \beta T-\frac{\pi}{3} \bigg)\notag\\
&\quad +\mu(\mu-1) \frac{\sqrt{3}}{2} \sin \bigg( \frac{\sqrt{3}}{2} \mu \beta T-\frac{\pi}{3} \bigg)\bigg\}\bigg]. \label{P1rfnonunit}
\end{align}
This also reduces to the ideal detection efficiency result (52) for $\eta=1$ ($\mu=0$). The unconditional distribution $P_1(T)$ does not vanish at $T=0$. Noting that the parameter $\mu$ increases from 0 to 1 as the efficiency of detection $\eta$ decreases from 1 to zero, Eqs. \eqref{w1rfnonunit} and \eqref{P1rfnonunit} then allow us to see that the effect of degrading detection efficiency is to damp out  oscillatory features and push the long-time tail of the distribution to be Poisson-like.  Simple expressions like this are not possible for other values of $\varOmega/\beta$. However, detailed calculations confirm that higher order distributions also follow this trend as detection efficiency decreases. Figure 8 illustrates the effect of non-unit detection efficiency on wait-time distributions ($n=1,2$) for $\varOmega/\beta=5\sqrt2$.  

As detection efficiency decreases, Rabi oscillations in both  $w_{n}(T)$ and $P_{n}(T)$ die out. For very small detection efficiency, $\eta \ll 1$, regardless of the strength of the field, the distributions for $w_{n}(T)$ and $P_{n}(T)$ are, qualitatively, very close to those for coherent light (dashed curves in Fig. 8), with the exception that $w_{1}(0)=0$.  For weak driving fields, photoemissions and, therefore, photo-detections (even with unit detection efficiency) become rare events. The wait-time distributions are then dominated by long wait-times, where they are  indistinguishable from the corresponding distributions for a Poisson photon sequence. Similarly, if the detection efficiency is very small, {\it photo-detections} become rare events (irrespective of photoemission rate or the strength of the driving field) and photo-detection wait-times are again dominated by long intervals pushing wait-time distribution close to those for random events.     

\subsection*{Mean and variance of wait-times}
The mean and variance of wait-times are also of interest as both can be measured experimentally.  The moments of $T$ with respect to $w_n(T)$  can be expressed in terms of its Laplace transform $\tilde{w}_{n}(s)$ as  
\begin{align}
\langle T^{m} \rangle_{w_{n}}= (-1)^{m}\frac{\partial^{m}}{\partial s^{m}} \tilde{w}_{n}(s) \bigg|_{s=0},
\end{align}
where  $\tilde{w}_{n}(s)$ is given by Eq.~\eqref{Lwnrf1}. The mean and variance of the wait-time with respect to $w_n(T)$ are then given by
  \begin{align}
\langle T \rangle_{w_n} &= \frac{n}{\eta \langle I \rangle_{ss}}, \label{avgTwn}\\
\langle (\Delta T)^{2}\rangle_{w_n} &= \frac{n}{(\eta \langle \hat{I} \rangle_{ss})^{2}} \bigg( 1- \frac{6\eta \beta^2 \Omega^{2}}{(\Omega^{2}+2\beta^{2})^{2}}\bigg), \label{avgDT2wn}
\end{align} 
where $\Delta T=T-\langle T\rangle$ and $\langle \hat{I} \rangle_{ss} = {\beta \Omega^2}/(\Omega^{2}+2\beta^{2})$ is the steady-state photon flux.

The averages of wait-time with respect to $P_{n}(T)$ can be obtained, similarly, or by using Eq. \eqref{Pnrf1} relating $P_n$ to $w_n$.  This leads to the following expressions for the first and second moments of wait-time with respect to $P_{n}(T)$ 
\begin{align}
\langle T \rangle_{P_n } &= \frac{n}{\eta \langle \hat{I} \rangle_{ss}} \bigg( 1-\frac{3 \eta \beta^2 \Omega^{2}}{n (\Omega^{2}+2\beta^{2})^{2}}\bigg), \label{avgTPn}\\
\langle T^{2} \rangle_{P_n } &= \frac{n(n+1)}{(\eta \langle \hat{I} \rangle_{ss})^{2}} + \frac{2}{\Omega^{2}+2\beta^{2}}\bigg(1-\frac{6 n \beta}{ \eta \langle \hat{I} \rangle_{ss}}\bigg).  \label{avgT2Pn}
\end{align}
Using these equations and Eq. \eqref{avgDT2wn}, the variance of unconditional wait-time can then be written as
\begin{align}
\langle (\Delta T) ^{2}\rangle_{P_n} = \langle (\Delta T) ^{2}\rangle_{w_{n}}+\frac{2\Omega^{2}-5\beta^2}{(\Omega^{2}+2\beta^{2})^{2}}. \label{avgDT2Pn}
\end{align}
A comparison of Eq. \eqref{avgDT2wn} with Eq. \eqref{tvarcoh} for Poisson photon sequence shows that the variance of wait-time between successive ($n=1$) photoemissions in resonance fluorescence is sub-Poissonian. In other words, the interval between successive photoemissions in resonance fluorescence fluctuates less than in a Poisson photon sequence, that is, successive photoemissions in resonance fluorescence are more {\it regular} than a Poisson photon sequence.  The smallest variance of photo-emission wait-times, $(1-3\eta/4)/(\eta\langle \hat{I}\rangle_{ss})^2$, occurs when $\varOmega/\beta=\sqrt2 $.  

For strong driving fields, ${\Omega^2}/{\beta^2} \gg 1$, the average photon flux saturates to $\langle \hat{I} \rangle_{ss} \equiv  {\beta \varOmega^2}/({\varOmega^{2}+2\beta^{2}})\to\beta$ or an average of one photoemission per 1/$\beta$ seconds.  In this limit, $ \langle T \rangle_{w_{n}}\approx \langle T \rangle_{P_{n}}\to n/\eta\langle \hat{I}\rangle_{ss}$,  which shows that for strong fields the average time for the $n$th photo-detection saturates to $n$ times the average interval $(\eta \beta)^{-1}$ between successive  photo-emissions and the variance approaches the Poisson limit $ \langle (\Delta T) ^{2}\rangle_{P_n} \approx  \langle (\Delta T) ^{2}\rangle_{w_{n}}\to n/\eta^2\beta^2$. 

\comment{The averages of wait-times with respect to $P_{n}(T)$ can be obtained by means of the formula
\begin{align}
\langle T^{m} \rangle_{P_{n}} &= \frac{m(m-1)}{(\Omega^{2}+2\beta^{2})} \langle T^{m-2} \rangle_{w_{n}}\nonumber \\ &-\frac{3\beta m}{(\Omega^{2}+2\beta^{2})} \langle T^{m-1} \rangle_{w_{n}}+\langle T^{m} \rangle_{w_{n}}, \label{avgTmPn}
\end{align}
which is obtained by using Eq. (50) relating $P_n$ to $w_n$ for resonance fluorescence.   Using this result, we find that the first and second moments of wait-time with respect to $P_{n}(T)$ are given by Eqs. (58) and (59). For  $\Omega = \beta$, $\langle \Delta T ^{2}\rangle_{w_{n}} =\frac{3n}{\eta^{2} \beta^{2}}(3-2\eta)$ showing that $\langle \Delta T ^{2}\rangle_{w_{n}} > \langle \Delta T ^{2}\rangle_{P_{n}}.$}
\section{Conclusions}
We have discussed the wait-time distributions for the $n$th photo-detections when a photo-electric detector is illuminated by a stationary light beam. The numbering of photo-detections -  1st, 2nd, 3rd, $\cdots$, begins after we start looking for photo-detections. If the counting begins at an arbitrary instant, the corresponding wait-time distributions are termed unconditional distributions. On the other hand, if counting begins only after a photo-detection is registered, that is, if the start of counting is conditioned on a photo-detection, the corresponding wait-time distributions are referred to as conditional distributions. These distributions characterize the temporal distribution of photo-detection events, which in turn result from the photon sequence underlying the incident light beam via the photoelectric effect. As the photon sequence itself is generated by photo-emission events at the source, the wait-time distributions characterize the temporal distribution of photons or time evolution of photoemissions. The wait-time distributions can be measured in photon counting experiments;  combined with the insights from photoelectron counting distribution and/or multi-time intensity correlations, they allow us to develop a physical picture of photon sequences generated by different sources. We also considered the effect of non-unit detection efficiency. When the efficiency of detection deviates from unity, photo-detection sequence (measured in the experiment) differs from the photon sequence incident on the detector or the source photo-emission sequence. For non-unit efficiency, we  may picture the photo-detection sequence as being obtained by random deletion of photons from the ideal photon sequence. This may begin to degrade certain distinctive and/or source specific features of the photon sequence. For sufficiently low detection efficiencies, the photo-detection sequence may become indistinguishable from that generated by a  Poisson photon sequence.    We demonstrated this by analyzing these distributions for three different sources of light and by comparing them with each other and with those for a Poisson photon sequence underlying coherent light.   

Thermal light (sometimes also called Gaussian light)  is an example of classical light.  Its statistical properties can be described in terms of an electric field amplitude that fluctuates as a complex Gaussian stochastic process. In this paper, we considered the source of thermal light to be a laser operating below threshold. The photon sequence underlying thermal light is thus generated by photoemissions from a laser cavity. This places thermal source on the same footing as the degenerate parametric oscillator below threshold and a coherently driven two-level atom, the latter two requiring quantum mechanics to describe their statistical properties. We present simple analytic expressions for the conditional and unconditional wait-time distributions, which are either exact or approximate but capture most significant features of the exact distributions.  These distributions carry distinctive features that can be traced to the dynamics of the source. For example, in the case of thermal light, there are two natural time scales in these distributions: $(2\gamma)^{-1}$, determined by the cavity lifetime and $(2\gamma\bar{n})^{-1}$, determined by the average photo-emission rate (photon flux). The former, prominent at low cavity photon number is quickly masked by the flux-limited wait-time as the mean cavity occupation number $\bar{n}$ increases.  The case of light from the DPO is especially interesting; cavity lifetime $(2\gamma)^{-1}$ appears in this case as well but at low mean cavity occupation number $\bar{n}$,  the second time scale is determined by the average flux of ``photon-pairs," $(\gamma\bar{n})^{-1}$ not by the average photon flux $2\gamma\bar{n}$.  In this small cavity occupation number regime,  photoemissions are rare events because the cavity is empty most of the time. The generation of a photon-pair inside the cavity suddenly boosts photo-emission rate lasting for a period of order $(2\gamma)^{-1}$. The average photo-emission rate remains small but has large fluctuations. This is reflected in the wait-time distributions, especially, $w_1(T)$, by the presence of the exponential term $e^{-2\gamma T}$[Eq. (35)]. It is also interesting that compared to thermal light the photo-emission rate is enhanced by a factor $1/\bar{n}$, which for small $\bar{n}$ can be a very large number. This enhancement is due to strong quantum correlations between the photons of a pair, which dominate photo-emission rate in the small $\bar{n}$ regime. We are also  able to see, analytically,  that with degradation of quantum efficiency, the wait-time distribution with distinctive time scales $(2\gamma)^{-1}$ and $\gamma\bar{n}$ approaches the distribution dominated by time scale $(2\gamma\bar{n})^{-1}$ determined by mean photon flux.  For large cavity occupation number $\bar{n}$, the wait-time distributions for the DPO are dominated by the time scale $(2\eta\gamma\bar{n})^{-1}$. The effect of detection efficiency is to simple scale the mean photo detection rate. Both of these features are similar to the behavior exhibited by thermal light.           

Wait-time distributions for resonance fluorescence have been considered before \cite{carmichael89,arnolduswn,arnolduswn2}. They are included here for completeness and comparison. Our contribution is to extend them to include the effect of non-unit quantum efficiency. Photon sequences in resonance fluorescence differ from those in thermal and DPO light in that the former exhibits anti-bunching while the latter two exhibit bunching in time.  This is reflected in the wait-time distributions, most notably, in the vanishing of $w_1(T)$, ruling out consecutive photo-detections with small wait-time. In contrast, $w_1(T)$  for thermal and DPO light has a maximum at $T=0$ indicating high probability of consecutive photo-detections with small wait-time.  This property is reflected in the variance of wait-times, which in resonance fluorescence is sub-Poissonian, that is, it is smaller than the wait-time variance for a Poisson photon sequence with the same photon flux. This is analogous to sub-Poissonian fluctuations of photocounts in a fixed interval $T$. For bunched  photon sequences from thermal and DPO sources, the variance of wait-times is super-Poissonian, that is, it exceeds the corresponding  variance for a Poisson photon sequence of the same flux from a coherent source. Once again, this is analogous to the super-Poissonian variance of photo-counts in an internal $T$.  The effect of non-unit detection efficiency is more complex in resonance fluorescence than it is for thermal light but in general,  decreasing detection efficiency wipes out distinctive features, such as Rabi oscillations, of wait-time distributions pushing photo-detection sequence closer to that generated by a Poisson photon sequence.  

Our analysis shows that unconditional and conditional wait-time distributions provide useful new insights for developing a physical picture of  photon sequences generated by different sources. We have seen that, for a given source, $P_1(T), P_2(T), w_1(T)$ and $w_2(T)$ provide the most  useful new information. They are also the most useful in comparing different sources. For example, the classification of photon sequences into sub-Poissonian, Poissonian, and super-Poissonian is based on these distributions. From a practical viewpoint also - they are relatively easier to measure and calculate.  Distributions beyond $n=3$ provide only incrementally new information. We also investigated the effect of detection efficiency on these distributions.  We find that the effect of non-unit detection efficiency is very different for classical and quantum light. For classical light, such as the thermal and coherent light,  the effect of  detection efficiency  is essentially to scale the effective  photo-detection rate, whereas for quantum light, such as the DPO far below threshold and two-level atom resonance fluorescence, it not only scales the effective photo-detection rate but also changes the shape of the distribution. In closing, we mention that the first member of the wait-time distribution family, $P_0(T)$, is the probability density that a time $T$ elapses without a photo-detection. As noted in the text following Eq. (3), this is really the generating function $G(1,T)$ for which we have the exact expression for all light sources considered in the paper. Interestingly, the measurements of $P_0(T)$ have been proposed as another way to quantify certain nonclassical features of light \cite{Bohmann}.   

With recent advances in photonic detector technology, high-efficiency detectors with high time and photon number resolution have become available and their use for more accurate measurements of photo-count probabilities and moments of light intensity for thermal light has been demonstrated \cite{cheng}. These quantities have been measured before even with low efficiency detectors.  Measurements of wait time distributions require  both high efficiency and high time and photon number resolution. These distributions describe a different aspect of photo-emissions - their temporal distribution.  In all cases considered in the paper, we are able to provide simple analytic expressions, which allow qualitative and quantitative insights into the temporal behavior of photons from different light sources.   We hope that the results presented in this paper will stimulate both theoretical calculations of wait-time distributions as well as their experimental measurements using the new high efficiency detectors. Such measurements will not only utilize the enhanced capabilities of the new detectors but also extend experimental measurements of photon statistics into a new direction to provide a more complete picture of photoemissions from light sources.

\clearpage 
\section*{Appendix: Wait-time distributions in resonance fluorescence}
In resonance fluorescence from a coherently driven two-level atom, it is more convenient to start with a calculation of $w_{n}(T)$ [Eq. \eqref{wnt1}] and then use it to calculate $P_{n}(T)$. Expanding the exponential in Eq.~\eqref{wnt1} and expressing the integrated flux operator $\hat{U}$ in terms of photon flux operator, $w_{n}(T)$ can be written as
\begin{widetext}
\begin{align}
w_{n}(T) &= \frac{1}{ \langle  \hat{I}(0) \rangle } \sum_{k=0}^{\infty} \frac{(k+n-1)!(-1)^{k}\eta^{n+k}}{(n-1)!k!}\times\notag\\
&\qquad \qquad \int_{0}^{T}dt_{k+n-1} \int_{0}^{t_{k+n-1} }dt_{k+n-2}\cdots \int_{0}^{t_{2} }dt_{1}  \langle \mathcal{T}:\hat{I}(T)\hat{I}(t_{k+n-1})\cdots\hat{I}(t_{1})\hat{I}(0)):\rangle,  \label{wnrf}
\end{align}
\end{widetext}
where we replaced $ \int_{0}^{T}dt_{k+n-1} \int_{0}^{T }dt_{k+n-2}\cdots \int_{0}^{T}dt_{1}$  by $(k+n-1)!\int_{0}^{T}dt_{k+n-1} \int_{0}^{t_{k+n-1} }dt_{k+n-2}\cdots\int_{0}^{t_{2} }dt_{1}$ \cite{lenstra}. The integrand of \eqref{wnrf} can be interpreted as the joint probability of detecting photons at the successive times $0,t_{1},t_{2},\cdots,t_{k+n-1},T$ \cite{carmichael89}. Using the Markov property of photoemissions, it can be written as a product of two-time conditional probabilities \cite{carmichael89,agarwal77}. To do so, we note that the probability of a photo-detection at $t_{k}$ conditioned upon a detection at $t_{k-1}(<t_{k})$ depends only on the interval $t_{k}-t_{k-1}$ and has the form
\begin{align}
\frac{\langle \mathcal{T}:\hat{I}(t_{k})\hat{I}(t_{k-1}):\rangle}{\langle \hat{I}(t_{k-1}) \rangle}=2\beta f_{0}(t_{k}-t_{k-1}),
\label{f0rf}
\end{align}
where $2 \beta$ is the Einstein-\textit{A} coefficient for the atomic transition and $f_0(t)$ describes atomic excitation when the atom starts initially in the ground  state \cite{singh83}. The integrand in Eq.~\eqref{wnrf} then can be written as (with $T = t_{n+k}$ and $t_0 =0$)
\begin{multline}
\langle \mathcal{T}:\hat{I}(T)\hat{I}(t_{k+n-1})\cdots\hat{I}(t_{1})\hat{I}(0)):\rangle = \\  \langle  \hat{I}(0) \rangle  \prod_{j=0}^{k+n-1}2\beta f_{0}(t_{j+1}-t_{j}).\label{rfproduct1}
\end{multline}
 Using Eqs.~\eqref{rfproduct1} in \eqref{wnrf} and taking the Laplace transform of the resultant expression, we obtain
\begin{align}
& \tilde{w}_{n}(s)= \sum_{k=0}^{\infty} \frac{(k+n-1)!(-1)^{k}}{(n-1)!k!} (2\beta \eta \tilde{f}_{0}(s))^{k+n},
\end{align}
where $\tilde{w}_{n}(s)$ and $\tilde{f}_{0}(s)$ are the Laplace transforms of $w_{n}(T)$ and $f_{0}(t_{k+1}-t_{k})$, respectively. The sum in the preceding equation can be recognized as a binomial series, so that $ \tilde{w}_{n}(s)$ can be written as
\begin{align}
& \tilde{w}_{n}(s)= \bigg( \frac{2\beta \eta \tilde{f}_{0}(s)}{1+2\beta \eta  \tilde{f}_{0}(s)} \bigg)^{n}.
\end{align}
Using the expression for $\tilde{f}_{0}(s)$ \cite{carmichael89,lenstra,arnolduswn}, 
\begin{equation}
\tilde{f}_{0}(s) = \frac{\Omega ^{2}}{2s((s+2\beta)(s+\beta)+\Omega ^{2})}, \label{lf0rf}
\end{equation}
where $\Omega$ is the Rabi frequency for the atomic transition, $\tilde{w}_{n}(s)$ can be written directly in terms of atomic parameters as
\begin{equation}
 \tilde{w}_{n}(s)= \bigg(\frac{\beta \eta  \Omega ^{2} }{s(s+\beta)(s+2 \beta)+\Omega ^{2}(s+\beta \eta )} \bigg)^{n}. \label{Lwnrf1}
\end{equation}

Using a similar procedure for $P_{n}(T)$, we  start with Eq.~\eqref{Pnt1} and express it in terms of intensity correlation functions, which factorize [Eq.  \eqref{rfproduct1}] 
\begin{align}
&\langle \mathcal{T}\textrm{:}\hat{I}(t_{n+k})\hat{I}(t_{n+k-1})...\hat{I}(t_{1})\textrm{:}\rangle\notag \\ &=(2\beta)^{n+k}f(t_1)\prod_{j=1}^{n+k-1}f_0(t_{j+1}-t_{j}),
\end{align}
where $f(t_1)$ is the probability of photoemission at $t_1$ when the atom starts in the steady state at $t=0$. We then find  \cite{lenstra}
\begin{align}
P_{n}(T)= &\sum_{k=0}^{\infty}\frac{(n+k-1)!(-1)^{k}(2\eta \beta)^{n+k}}{(n-1)!k!}\times\notag \\ &f(t_1)\prod_{j=1}^{n+k-1}f_0(t_{j+1}-t_{j}).
\end{align}
Taking the Laplace transform of $P_n(T)$ and using the Laplace transform  of $f_0(T)$ 
\comment{
Laplace transform of this expression yields 
\begin{align}
&\tilde{P}_{n}(s)= 2\eta \beta \tilde{f}(s)\sum_{k=0}^{\infty}\frac{(-1)^{k}(n+k-1)!}{(n-1)!k!}(2\eta \beta\tilde{f}_{0}(s))^{n+k-1},
\end{align}
}
given by Eq.~\eqref{lf0rf} and of $f(t_1)$ given by
\begin{align}
\tilde{f}(s) = \frac{\Omega^{2}}{2s(\Omega^{2}+2\beta^2)}, 
\end{align}
 we get 
\begin{align}
&\tilde{P}_{n}(s)= 2\eta \beta \tilde{f}(s)\sum_{k=0}^{\infty}\frac{(-1)^{k}(n+k-1)!}{(n-1)!k!}(2\eta \beta\tilde{f}_{0}(s))^{n+k-1}.
\end{align}
The sum can be carried out to yield  
\begin{align}\label{LaplacePnrf}
 \tilde{P}_{n}(s)&= \bigg( \frac{s(s+3\beta)}{\Omega^{2}+2\beta^2}+1 \bigg)\bigg(\frac{2\eta \beta \tilde{f}_{0}(s)}{1+2\eta \beta \tilde{f}_{0}(s) }\bigg)^{n}\,, \notag
 \\
&\equiv (C s^{2}+3\beta C s+1) \tilde{w}_{n}(s),
\end{align}
 where $C = \frac{1}{\Omega^{2}+2\beta^2}$. Using this expression and the properties of the Laplace transform \cite{arfken}, we find that  $P_{n}(T)$ can be expressed in terms of  $w_{n}(T)$ as
  \begin{align}
P_{n}(T)&= C\frac{d^2}{dT^{2}}w_{n}(T)+3\beta C \frac{d}{dT}w_{n}(T)+w_{n}(T)\,. \label{Pnrf1}
\end{align} 

The conditional distribution $w_n(T)$ is found by taking the inverse Laplace transform of $\tilde{w}_{n}(s)$ [Eq.~\eqref{Lwnrf1}]  using the calculus of residues \cite{kimble76,lenstra,carmichael89}. This leads to the following expression for  $w_n(T)$ 
\begin{widetext}
\begin{flalign}
&w_{n}(T)=\frac{\beta(\Omega/\beta)^{2n}e^{-\beta T}}{((n-1)!)^{3}(1-\Omega ^{2}/\beta^2)^{n}}\sum_{k=0}^{n-1}(\beta T)^{n-k-1}\left[(-1)^{n}\mathcal{D}_{0}(n,k)+\mathcal{D}(n,k)((-1)^{k}e^{\sqrt{\beta^{2}-\Omega ^{2}}T} +e^{-\sqrt{\beta^{2}-\Omega ^{2}}T})\right],& \label{wnrf1}\\
& \text{where}\quad \mathcal{D}_{0}(n,k) = {n-1 \choose k}\frac{1}{(\sqrt{1-\Omega ^{2}/\beta^{2}})^{k}}  \sum_{j=0}^{k}(-1)^{j}{k \choose j} (n+k-j-1)!(n+j-1)! \,,\label{D0}&\\
&\text{and}\qquad
 \mathcal{D}(n,k) = {n-1 \choose k}\frac{1}{(\sqrt{1-\Omega ^{2}/\beta^{2}})^{k}}  \sum_{j=0}^{k}{k \choose j} \frac{(n+k-j-1)!(n+j-1)!}{2^{n+j}}. & \label{D}
\end{flalign}
\end{widetext}
\comment{\begin{widetext}
\begin{flalign}
&w_{n}(T)=\frac{(\beta \Omega^{2})^{n}}{((n-1)!)^{3}(\beta^{2}-\Omega ^{2})^{n}}e^{-\beta T}\sum_{k=0}^{n-1}T^{n-k-1}\left[(-1)^{n}\mathcal{D}_{0}(n,k)+\mathcal{D}(n,k)((-1)^{k}e^{\sqrt{\beta^{2}-\Omega ^{2}}T} +e^{-\sqrt{\beta^{2}-\Omega ^{2}}T})\right],& \label{wnrf1}\\
& \text{where}\quad \mathcal{D}_{0}(n,k) = {n-1 \choose k}\frac{1}{(\sqrt{\beta^{2}-\Omega ^{2}})^{k}}  \sum_{j=0}^{k}(-1)^{j}{k \choose j} (n+k-j-1)!(n+j-1)! \,,\label{D0}&\\
&\text{and}\qquad
 \mathcal{D}(n,k) = {n-1 \choose k}\frac{1}{(\sqrt{\beta^{2}-\Omega ^{2}})^{k}}  \sum_{j=0}^{k}{k \choose j} \frac{(n+k-j-1)!(n+j-1)!}{2^{n+j}}. & \label{D}
\end{flalign}
\end{widetext}}
In general, these expressions must be evaluated numerically. Nevertheless, it can be seen from these equations that $w_{n}(0)=0$ and $w'_{n}(T)=0$. The former is a reflection of the fact that a two-level atom can emit only one photon at a time. 
 The conditional wait-time distributions for $n=1-3$ are then found, from Eqs.~\eqref{wnrf1}-\eqref{D}, to be those given by Eqs. (52) -- (54). The unconditional wait-time distribution $P_{n}(T)$ can be obtained by taking the inverse Laplace transform of Eq.~\eqref{LaplacePnrf} or by using the formula for $w_n(T)$ in Eq. \eqref{Pnrf1}.  

\subsection*{Nonunit detection efficiency} 
For non-unit detection efficiency, the expressions for $w_n(T)$ and $P_n(T)$ become even more cumbersome. However, by  introducing some auxiliary quantities, they can be written in a form similar to Eqs. \eqref{wnrf1}-\eqref{D}.  The cubic $s(s+\beta)(s+2\beta)+\Omega^2(s+\eta \beta)$ in the denominator of Eq. (\ref{Lwnrf1}) can be factored as  $(s-s_1)(s-s_2)(s-s_3)$, where $s_1, s_2,$ and $s_3$ are given by \cite{abramowitz}
\begin{subequations}
\begin{align}
 s_1 &=\beta( -1 +\delta_{1}+\delta_{2}),\\
   s_2 &= \beta(-1-\delta_{1}e^{i\pi/3}-\delta_{2}e^{-i\pi/3}) \\
s_3 &=\beta(-1-\delta_{1}e^{-i\pi/3}-\delta_{2}e^{i\pi/3})\,,
\comment{\begin{align}
 s_1 &= -\beta +\Delta_{1}+\Delta_{2},\\
 s_2 &= -\beta-\Delta_{1}e^{i\pi/3}-\Delta_{2}e^{-i\pi/3} \\
s_3 &=-\beta-\Delta_{1}e^{-i\pi/3}-\Delta_{2}e^{i\pi/3}\,,}
\end{align}
with 
\end{subequations}
\begin{subequations}\label{w1rfnonunitroots} \begin{align}
\delta_1&= \frac{\sqrt[3]{2}(1-\varOmega^2/\beta^2)}{B},\quad \delta_2=\frac{B}{3\sqrt[3]{2}},\\
B &=\Big[27( 1-\eta)(\varOmega^2/\beta^2)\notag\\
&+\sqrt{108\left(\frac{\Omega^2}{\beta^2}-1\right)^3+\left(27( 1-\eta)\frac{\Omega^2}{\beta^2}\right)^2}\Big]^{1/3}.
\comment{ \Delta_1&= \frac{\sqrt[3]{2}\omega^2}{B},\quad \Delta_2=\frac{B}{3\sqrt[3]{2}},\\
B &=\Big[27\beta( 1-\eta)\Omega^2\notag\\
&+\sqrt{108(\Omega^2-\beta^2)^3+(27\beta( 1-\eta)\Omega^2)^2}\Big]^{1/3}\,.}\end{align}
\end{subequations}
As a check, for $\eta = 1$,  $B \to  \sqrt[3]{2}\sqrt{3(\Omega^2/\beta^2-1)}$ and we recover $
s_1 = -\beta ,\, s_2 = -\beta-\sqrt{\beta^2-\Omega^2} $ and $ s_3 = -\beta+\sqrt{\beta^2-\Omega^2}$.   In terms of the quantities introduced in Eqs. (78a) and (78b), the nonunit detection efficiency  expressions for $w_{n}(T)$ then can be written in a form similar to Eq. \eqref{wnrf1}:
\begin{widetext}
 \begin{equation}
w_{n}(T)=C_{n}e^{-\beta T} \sum_{k=0}^{n-1} \binom{n-1}{k}\frac{(\beta T)^{n-1-k}}{(\sqrt{3(\delta_{1}^{2}+\delta_{1}\delta_{2}+\delta_{2}^{2})})^{k}}((-1)^{k}\mathcal{J}_{0}(n,k)e^{(\delta_1 +\delta_2)\beta T}+2(-1)^{n}\mathcal{J}(n,k,T)e^{- (\delta_{1}+\delta_{2})\beta T/2}),
\end{equation}
where 
\begin{equation}
\begin{aligned}
&
 C_{n}= \frac{(\eta \Omega^2/\beta^2)^{n}}{((n-1)!)^3 3^n(\delta_{1}^{2}+\delta_{1}\delta_{2}+\delta_{2}^{2})^{n}}, \\
& \mathcal{J}_{0}(n,k) =\sum_{p=0}^{k} \binom{k}{p}(n+k-p-1)!(n+p-1)!\cos((2p- k)\theta_1),\\
& \mathcal{J}(n,k) = (\sqrt{3})\bigg(\frac{\sqrt{\delta_{1}^{2}+\delta_{1}\delta_{2}+\delta_{2}^{2}}}{\delta_2-\delta_1} \bigg)^{n}\times \\ &\sum_{p=0}^{k} \binom{k}{p}(-1)^{p}(n+k-p-1)!(n+p-1)!\bigg(\frac{\sqrt{\delta_{1}^{2}+\delta_{1}\delta_{2}+\delta_{2}^{2}}}{\delta_2-\delta_1} \bigg)^{p} \cos \bigg(\frac{\delta_2-\delta_1}{2}\sqrt{3}\beta T+\phi_{nkp}\bigg), \\
&\theta_{1}= \arctan\bigg( \frac{1}{\sqrt{3}} \frac{\delta_2-\delta_1}{\delta_2+\delta_1}\bigg) \; \; \textrm{and}\; \;
\phi_{nkp} = \theta_1(n+k-p)-(n+p)\frac{\pi}{2}.
\end{aligned}
\end{equation}
\comment{
 \begin{equation}
w_{n}(T)=C_{n}e^{-\beta T} \sum_{k=0}^{n-1} \binom{n-1}{k}\frac{T^{n-1-k}}{(\sqrt{3(\Delta_{1}^{2}+\Delta_{1}\Delta_{2}+\Delta_{2}^{2})})^{k}}((-1)^{k}\mathcal{J}_{0}(n,k)e^{\Delta_1 T+\Delta_2 T}+2(-1)^{n}\mathcal{J}(n,k)e^{-\frac{\Delta_{1}+\Delta_{2}}{2}T}),
\end{equation}
where 
\begin{equation}
\begin{aligned}
&\omega = \sqrt{\beta^2-\Omega^{2}},\; \;
 C_{n}= \frac{(\eta \beta \Omega^2)^{n}}{((n-1)!)^3 3^n(\Delta_{1}^{2}+\Delta_{1}\Delta_{2}+\Delta_{2}^{2})^{n}}, \\
& \mathcal{J}_{0}(n,k) =\sum_{p=0}^{k} \binom{k}{p}(n+k-p-1)!(n+p-1)!\cos((2p- k)\theta_1),\\
& \mathcal{J}(n,k,T) = (\sqrt{3})\bigg(\frac{\sqrt{\Delta_{1}^{2}+\Delta_{1}\Delta_{2}+\Delta_{2}^{2}}}{\Delta_2-\Delta_1} \bigg)^{n}\times \\ &\sum_{p=0}^{k} \binom{k}{p}(-1)^{p}(n+k-p-1)!(n+p-1)!\bigg(\frac{\sqrt{\Delta_{1}^{2}+\Delta_{1}\Delta_{2}+\Delta_{2}^{2}}}{\Delta_2-\Delta_1} \bigg)^{p} \cos \bigg(\frac{\Delta_2-\Delta_1}{2}\sqrt{3}T+\phi_{nkp}\bigg), \\
&\theta_{1}= \arctan\bigg( \frac{1}{\sqrt{3}} \frac{\Delta_2-\Delta_1}{\Delta_2+\Delta_1}\bigg) \; \; \textrm{and}\; \;
\phi_{nkp} = \theta_1(n+k-p)-(n+p)\frac{\pi}{2}.
\end{aligned}
\end{equation}}
\end{widetext}
Like Eq. \eqref{wnrf1}, in general, this must be evaluated numerically. For $n=1$ and $\varOmega/\beta=1$, the expressions for  $w_1(T)$  and $P_1(T)$ [using $w_1(T)$ in Eq. \eqref{Pnrf1}] simplify.  These are given by  Eqs. \eqref{w1rfnonunit} and \eqref{P1rfnonunit}, respectively, in the main paper.



\begin{thebibliography}{99}

\bibitem{mandelwolf65} L. Mandel and E. Wolf,{\it  Coherence Properties of Optical Fields},  Rev. Mod. Phys. {\bf 37} 231-287 (1965).

\bibitem{bendjaballah}
C. Bendjaballah and  F. Perrot, \textit{Statistical properties of intensity‐modulated coherent radiation. Theoretical and experimental aspects.} J. Appl. Phys. \textbf{44}, 5130 (1973).

\bibitem{saleh} B.E.A. Saleh, {\sl Photoelectron Statistics} (Springer-Verlag, Berlin, 1978). 

\bibitem{carmichael89}
H.J. Carmichael, S. Singh, R. Vyas and P.R. Rice, \textit{Photoelectron waiting times and atomic state reduction in resonance fluorescence}, Phys. Rev. A\textbf{39}, 1200 -1217 (1989).

 \bibitem{glauber63}
R.J. Glauber, {\sl The Quantum Theory of Optical Coherence}, Phys. Rev. {\bf 130}, 2529-2539 (1963). 

\bibitem{vyassingh88}
R. Vyas and S. Singh, {\sl Waiting-time distributions in the photodetection of squeezed light}, Phys. Rev. A{\bf 38}, 2423-2430 (1988).

\bibitem{vyassingh00} 
 R. Vyas and S. Singh. \textit{ Antibunching and photoemission waiting times}. J. Opt. Soc. Am. B{\bf 17}, 634-637 (2000).

\bibitem{sipahigil} A. Sipahigil, M. L. Goldman, E. Togan, Y. Chu, M. Markham, D. J. Twitchen, A. S. Zibrov, A. Kubanek,
and M. D. Lukin, {\it Quantum interference of single photons
from remote nitrogen-vacancy centers in diamond}, Phy. Rev. Lett. {\bf 108}, 143601 (2012).

\bibitem{leifgen} M. Leifgen, T. Schr\"{o}der, F. G\"{a}deke, R. Riemann,
V. M\'{e}tillon, E. Neu, C. Hepp, C. Arend, C. Becher,
K. Lauritsen, and Oliver Benson {\it Evaluation of nitrogen- and silicon-vacancy defect centres as single photon sources in quantum key distribution}, New J. Phys. {\bf 16}, 023021 (2014).

\bibitem{couteau} Christophe Couteau, Stefanie Barz, Thomas Durt, Thomas Gerrits, Jan Huwer, Robert Prevedel, John Rarity, Andrew Shields, and  Gregor Weihs, {\it  Applications of single photons to quantum communication and computing}, Nature Reviews Physics {\bf 5}, 326--338 (2023).

\bibitem{cheng} 
Risheng Cheng, Yiyu Zhou, Sihao Wang, Mohan Shen, Towsif Taher, and Hong X. Tang, {\it A 100-pixel photon-number-resolving detector unveiling photon statistics}, Nature Photonics {\bf 17}, 112--119 (2023).

\bibitem{haken70} H. Haken, {\it Laser Theory} (Springer-Verlag, New York, 1984). Section VI.4. 
 
\bibitem{loudon00} R. Loudon, {\it The quantum theory of light} (Oxford University Press, Oxford, UK, 2000). Chapter 3. 

 \bibitem{microcavity} F. Brange, P. Menczel, and C. Flindt, {\sl Photon counting statistics of a microwave cavity},  Phys. Rev. B{\bf 99}, 085418 (2019)

\bibitem{drummond} P.D. Drummond, K.J. McNeil, and D.F. Walls,{\sl Non-equilibrium transitions in sub/second harmonic generation}  Opt. Acta {\bf 28}, 211-225 (1981).

\bibitem{wolinsky} H.J. Carmichael and M.J. Wolinsky, {\it Quantum noise in the parametric oscillator: From squeezed states to coherent-state superpositions}, Phys. Rev. Lett. {\bf 60}, 1836-1839 (1988). 

\bibitem{kimble87} H.J. Kimble and D.F. Walls (Guest eds.) {\sl Special issue  on Squeezed States of the Electromagnetic Feld}, J. Opt. Soc. Am. B {\bf 4} (10) (1987). 

\bibitem{vyas95} R. Vyas and  S. Singh, \textit{Exact Quantum Distribution for Parametric Oscillators}, Phys. Rev. Lett. \textbf{74}, 2208--2211 (1995).

\bibitem{vyas03} S. Siddiqui, D. Erenso, R. Vyas, and S. Singh, {\it Conditional measurements as probes of quantum dynamics}, Phys. Rev. A{\bf 67}, 063808 (2003). 

\bibitem{vines06} 
J. Vines, R. Vyas, and S. Singh, \textit{Conditional homodyne detection of light with squeezed quadrature fluctuations}, Phys. Rev. A\textbf{74}, 023817 (2006).

\bibitem{vyas09}  R. Vyas and  S. Singh, \textit{Statistical properties of light from optical parametric oscillators}, Phys. Rev. A\textbf{80}, 063836 (2009).

\bibitem{arnab10} 
A. Mitra, R. Vyas, and S. Singh,  \textit{Nonclassicality of light from a degenerate parametric oscillator}. J. Mod. Opt. {\bf 57}, 1293-1299 (2010).

\bibitem{christ13} A. Christ, A. Fedrizzi, H. Hubel, T. Jennewein, C. Silberhorn, {\it Parametric Down-Conversion} in {\it  Single-Photon Generation and Detection Physics and Applications}, eds. A. Migdall, S.V. Polyakov, J. Fan, and J.C. Bienfang, (Academic Press, 2013). Chapter 11.

\bibitem{scullyzubairy} M.O. Scully and M.S. Zubairy, {\it Quantum Optics} (Cambridge University Press, New York, 1997). Chapter 11.

\bibitem{carmichael76} H.J. Carmichael and D.F. Walls, {\it A quantum-mechanical master equation treatment of the dynamical Stark effect}, J. Phys. B: Atom. Mol. Phys. {\bf 9}, 1199-1219 (1976).

\bibitem{kimble76} H.J. Kimble and L. Mandel, {\it Theory of Resonance fluorescence}, Phys. Rev. A{\bf 13}, 2123 - 2144 (1976). 

\bibitem{arnoldus} 
H.F. Arnoldus and G. Nienhuis, \textit{Photon statistics of fluorescence radiation}. Opt. Acta {\bf 33}, 691-702 (1986). 
 
 \bibitem{arnolduswn}
H.F. Arnoldus and R.A. Rielhe, \textit{Waiting times, probabilities and the $Q$ factor of fluorescent photons}, J. Mod. Opt. {\bf 59}, 1002-1015 (2012).

\bibitem{arnolduswn2}
H.F. Arnoldus and R.A. Rielhe. \textit{Conditional probability densities for photon emission in resonance fluorescence}, Phys. Lett. A\textbf{376}, 2584-2587 (2012).

\bibitem{arfken}
 G.B. Arfken, H. Weber, and  F.E. Harris. \textit{ Mathematical methods for physicists: A comprehensive guide}. (Academic Press, Oxford, 2013). Chapter 20. 

\bibitem{bedard} G. Bedard, {\sl Photon Counting Statistics of Gaussian Light},  Phys. Rev. {\bf 151}, 1038-1039 (1966).

\bibitem{jakemanandpike}
E. Jakeman and E.R. Pike, {\it The intensity-fluctuation distribution of Gaussian light}, J. Phys. A: Gen. Phys. {\bf 1}, 128-138 (1968).  

\bibitem{luis23} Luis Felipe Morales Bultron, \textit{Photon Counting Statistics of Classical and Quantum Light Sources} (MS Thesis).  University of Arkansas, Fayetteville, AR (2023)

\bibitem{vyassingh89-1} 
R. Vyas and  S. Singh, \textit{Photon-counting statistics of the degenerate optical parametric oscillator}. Phys. Rev. A\textbf{40}, 5147-5159 (1989). 

 \bibitem{vyassingh89-2}
 R. Vyas and S. Singh, {\it Quantum statistics of broadband squeezed light}, Opt. Lett. {\bf 14}, 1110-1112 (1989).
 
 \bibitem{huang} J. Huang and P. Kumar,  {\sl Photon-counting statistics of multimode squeezed light}, Phys. Rev. A{\bf 40}, 1670 -1673 (1989).

 \bibitem{friberg85} S. Friberg, C.K. Hong, and L. Mandel, {\it  Intensity dependence of the normalized intensity correlation function in parametric down-conversion}, Opt. Commun. {\bf 54}, 311-316 (1985).
 
\bibitem{carmichael93}
H.J. Carmichael, \textit{An Open systems Approach to Quantum Optics}. (Springer-Verlag, Berlin, 1993).

\bibitem{mandel77} L. Mandel, {\sl Sub-Poissonian photon statistics in resonance fluorescence}, Opt. Lett. {\bf 4}, 205-207 (1977).

\bibitem{lenstra} 
D. Lenstra,  \textit{Photon-number statistics in resonance fluorescence}, Phys. Rev. A{\bf 26}, 3369-3377 (1982). 

 \bibitem{singh83} S. Singh, {\it Antibunching, sub-Poissonian photon statistics and finite bandwidth effects in resonance fluorescence}, Opt. Commun. {\bf 44}, 254-258 (1983).  

\bibitem{agarwal77} G. S. Agarwal, {\it Time factorization of the higher-order intensity correlation functions in the theory of resonance fluorescence},  Phys. Rev. A{\bf 15}, 814-816 (1977)

\bibitem{abramowitz}
M. Abramowitz and I.A. Stegun, \textit{Handbook of Mathematical Functions } (Dover, New York, 1965). Sec. 3.8.2. 

\bibitem{Bohmann} Martin Bohmann and Elizabeth Agudelo, {\it Phase-Space Inequalities Beyond Negativities}, Phys. Rev. Lett. {\bf 124}, 133601 (2020). 
\end{thebibliography}
\end{document}